\documentclass[11pt,a4paper,nofootinbib]{article}
\pdfoutput=1
\usepackage{jheppub}
\usepackage[utf8]{inputenc}
\usepackage{graphicx,floatflt,amssymb,rotate, cancel}
\usepackage{mathrsfs}
\usepackage{amssymb}
\usepackage{amsmath}
\usepackage{bm}
\usepackage{color}
\usepackage{xcolor}
\usepackage{graphicx}
\usepackage{multirow}
\usepackage{array}
\usepackage{float}
\usepackage[toc,page]{appendix}
\usepackage{diagbox}
\usepackage[mathscr]{euscript}
\usepackage{cleveref}
\usepackage{soul}

\usepackage{subfig}
\usepackage{hyperref}
\usepackage{url}

\makeatletter
\gdef\@fpheader{}
\makeatother

\definecolor{orange}{rgb}{1,0.5,0}

\newcommand{\met}{\ensuremath{/ \hspace{-.7em} E_T}}

\title{Searching for exotic Higgs bosons at the LHC}
\author[a,d)]{Gautam Bhattacharyya,}
\author[b,e)]{Siddharth Dwivedi,}
\author[c)]{Dilip Kumar Ghosh,} 
\author[a,d)]{Gourab Saha,} 
\author[a,d)]{Subir Sarkar}
\affiliation[a)]{Saha Institute of Nuclear Physics, 1/AF Bidhan Nagar,
  Kolkata 700064, India}
\affiliation[b)]{Institute of Physics,
P.O. Sainik School, Sachivalaya Marg, Bhubaneswar
751005, India}
\affiliation[c)]{School of Physical Sciences, Indian Association for
  the Cultivation of Science,
2A $\&$ 2B, Raja S.C. Mullick Road, Kolkata 700032, India}
\affiliation[d)]{Homi Bhabha National Institute, Training School
  Complex, Anushaktinagar, Mumbai 400094, India}
\affiliation[e)]{Krea University, 5655, Central Expressway, Sri City,
  Andhra Pradesh, 517646, India}
\emailAdd{gautam.bhattacharyya@saha.ac.in}
\emailAdd{siddharth.d@iopb.res.in}
\emailAdd{tpdkg@iacs.res.in}
\emailAdd{gourab.saha@saha.ac.in}
\emailAdd{subir.sarkar@cern.ch}

\abstract{We analyse in a model independent way the possibilities of
  digging out neutral exotic Higgs states, should they exist endowed
  with unconventional couplings with ordinary matter and gauge fields,
  at the 14 TeV run of the Large Hadron Collider (LHC), adding some
  comparative studies for 13.6 and 13 TeV runs. Flavor models, based
  on some discrete symmetry groups, with extended scalar sectors are
  known to yield exotic spin-$0$ states, both CP-even and CP-odd, with
  purely flavor off-diagonal Yukawa couplings. The gauge interaction
  of one such CP-even state is also unusual that, unlike the Standard
  Model Higgs boson, it does not couple to gauge boson pairs. Such
  unconventional properties immune these exotic states from receiving
  traditional collider and electroweak constraints, and hence those
  states could be light.  Without committing to any specific model,
  exploiting their peculiar Yukawa and gauge properties, we explore
  the discovery potential of those exotic Higgs states through some
  interesting topologies by figuring out some specific kinematic
  variables that suppress the backgrounds.}

\begin{document}
\maketitle

\section{Introduction} \label{intro}

Ever since the ATLAS and CMS Collaborations of the CERN LHC discovered
the 125 GeV Higgs boson, thus completing the particle spectrum of the
Standard Model\,(SM), the hunters of physics beyond the SM\,(BSM) have
intensified their searches for any other Higgs-like boson(s).  Indeed,
there are motivations to hypothesize an underlying extended scalar
sector. One of them is the r\^ole of additional scalars in
facilitating explanation to the flavor problem. Specifically, discrete
flavor symmetries have been successfully employed to explain the quark
and lepton masses and mixing \cite{Altarelli:2010gt, Ma:2007ia,
  Ishimori:2010au}.  The byproducts are no less interesting
either. With enlarged scalar spectra, many of these flavor models
contain exotic spin-$0$ states endowed with apparently weird couplings
to fermions and gauge bosons. Exploiting those unconventional
couplings, how to dig those exotic scalar\,(pseudoscalar) states out
of the debris of the 14 TeV LHC is the subject matter of the present
paper.

Although our approach is sufficiently model independent, to set up the
context, we start our discussion with a reference to a class of flavor
models based on the group $S_3$ introduced in
\cite{Pakvasa:1977in}. These models contain enlarged scalar sectors
with nonstandard couplings to fermions and gauge bosons. $S_3$ is the
smallest non-Abelian discrete group generating the symmetry of an
equilateral triangle. It has two singlet\,($\underline{1},
\underline{1}'$) and one doublet\,($\underline{2}$) irreducible
representations. The doublet representation facilitates maximal
mixing, and together with the two inequivalent singlets, $S_3$ can
satisfactorily reproduce the fermion masses and mixing. However, what
plays a crucial role in these explanations is the presence of three
copies of SU(2) doublet Higgs bosons, $\phi_{1,2,3}$, out of which
$\phi_{(1,2)}$ form an $S_3$ doublet and $\phi_3$ remains a singlet. A
rich scalar spectrum emerges, with three CP-even and two CP-odd
neutral, plus two sets of charged scalars. The details of the
minimization of the potential, mass spectra of the
scalars\,(pseudoscalars) and their couplings to the gauge and matter
fields may be found in~\cite{Bhattacharyya:2012ze,
  Bhattacharyya:2010hp}. One of the CP-even neutral scalars turns out
to be the SM Higgs, which we denote by $h_{125}$. Of the nonstandard
states, except a CP-even state ($H$) and a CP-odd state ($\chi$), the
rest may be considered to be sufficiently heavy having couplings to
the gauge and matter fields resembling those in the two-Higgs doublet
models. With this background, we may forget any model specific details
of the scalar spectrum, except concentrating on two peculiar
properties of $H$ and $\chi$, which we shall discuss shortly.  We also
point out that $\Delta(27)$, as the smallest group which provides a
source of geometric CP violation, also contains a scalar and
pseudoscalar having similar properties~\cite{Bhattacharyya:2012pi},
see also~\cite{Luhn:2007uq,Ishimori:2010au}. In this paper we
investigate, for the first time, how one can exploit the peculiar
behavior of $H$ and $\chi$ towards the gauge and matter fields to
detect those exotic states at the upcoming 14 TeV run of the LHC.

To propel our discussion this far, we had to draw inspiration from the
flavor models. From now onward, we do not appeal to any specific
model, as we know that a large class of well-motivated flavor models,
which contain three Higgs doublets, carry such exotic
scalar\,(pseudoscalar) states. In fact, from the point of view of an
unbiased experimental searches, we merely assume that some underlying
scalar sector, regardless of its origin, gives us $H$ and $\chi$ with
the following nonstandard properties:

\begin{itemize}

\item There are no $H VV$-type couplings, where $V \equiv W^\pm,
  Z$. The $H \chi Z$ coupling takes the simple form\,($q_{\mu} \equiv$
  momentum transfer): $$H \chi Z~:~
  \left(\frac{-ie}{2\sin\theta_W\cos\theta_W}\right) q_\mu$$
\item $H (\chi)$ has {\em only} flavor off-diagonal Yukawa
  couplings. The relevant piece of Yukawa Lagrangian is $$Y_{ff'}
  \bar{f}\,(i\gamma^5)\,f' H\,(\chi) ~+ ~ {\rm h.c.}$$ To be more
  specific, $f,f' \equiv e\mu, \mu\tau, e\tau, uc, tc, ut, ds, db,
  sb$.
  
\end{itemize} 

Because there is no $HVV$ coupling, neither the LEP2 limit nor the
electroweak precision constraints would apply on the mass of $H$. The
pseudoscalar $\chi$ does not couple to $VV$ anyway.  Moreover, since
neither $H$ nor $\chi$ has any diagonal Yukawa coupling, the usual LHC
constraints do not apply on their masses either. Therefore, both $H$
and $\chi$ could be light~\cite{Bhattacharyya:2012ze}. Since the
choices of their masses would greatly influence the search strategies,
we cannot but make a few working assumptions before we start our
analyses. We first select a few benchmark points assuming $m_{H} \approx m_t$, in a way that the top quark
does not have a sizable branching fraction into $H$ and a charm
quark. Later we extend the range of $m_H$ mostly to the higher side.
We assume $\chi$ to be sufficiently light, at least much
lighter than $H$.  Indeed, any other choices could be equally likely,
but the possibility of a not-so-heavy exotic scalar and a lighter
pseudoscalar is in conformity with the lore and excitement prevailing
in the community for a while. Similarly, the size of the $H/ \chi$
off-diagonal Yukawa couplings would impact the search strategies.

For any specific flavor symmetry group, those purely off-diagonal
Yukawa couplings have a r\^ole to play in reproducing the fermion
masses and mixing. Studies with the $S_3$ group have shown that in $H
qq'$ coupling, one of $q$ and $q'$ has to be necessarily a third
generation quark~\cite{Bhattacharyya:2010hp, Bhattacharyya:2012ze}. In
the present analysis we remain agnostic about this requirement and
treat the generations democratically from the perspective of model
blind experimental searches.  However, we do keep in mind the
extremely tight constraints on $H ds$ and $\chi ds$ couplings from
$K_L \to \mu e$ decays, specially when we are dealing with $H$ and
$\chi$ masses of ${\cal O}$ ($10\,-\,100$)
GeV~\cite{ParticleDataGroup:2020ssz}.  To circumvent this, we set the
off-diagonal $Hds$ and $\chi ds$ couplings to zero. We now pay
attention to the $Huc$ and $\chi uc$ Yukawa couplings ($Y_q^H$ and
$Y_q^\chi$, respectively), each of which will contribute to the $D^0 -
\bar{D}^0$ mixing.  Although the constraints from this mixing are not
as tight as from $K^0 - \bar{K}^0$ or $B_d^0 - \bar{B_d}^0$ mixing,
still for ${\cal{O}}$\,($100$) GeV mediator masses the upper limit on
the corresponding Yukawa couplings would be roughly $10^{-4}$. Now,
tree level meson mixing amplitute goes as $(\frac{{Y_q^H}^2}{m_H^2} -
\frac{{Y_q^\chi}^2}{m_{\chi}^2})$, i.e. a scalar and a pseudoscalar
contribute with opposite sign (see e.g. \cite{Botella:2015hoa}).  In
the present analysis, we set $Y_q^\chi \approx \frac{m_\chi}{m_H}
Y_q^H$, to relax the above stringent constraint from $D^0 - \bar{D^0}$
mixing by one or two orders of magnitude.  We arrange for this partial
cancellation to take advantage of our model that contains not only a
light scalar but simultaneously a light pseudoscalar, both having
purely off-diagonal Yukawa couplings. It is important to note that
$Y_q^H$ and $Y_q^{\chi}$ need not be strictly tuned to drive home the
essential features of our analysis. These couplings will play a
significant r\^ole in the production of these exotic spin-$0$ states
at the LHC. The off-diagonal couplings involving the top quark would
not be so relevant for our analysis.

On the leptonic sector, nonobservation of various lepton flavor
violating\,(LFV) processes, like $\ell_i \to \ell_j \gamma $ (with
$\ell_i \equiv \tau, \mu $ and $\ell_j \equiv \mu, e$), $\mu + N \to e
+N $ (i.e. $\mu-e$ conversion)~\cite{Lindner:2016bgg}, as well as
$e^+e^- \to \mu^+\mu^- (\tau^+ \tau^-)$ put a very strong limit on the
product of LFV Yukawa couplings involving the first two generations
$(e, \mu)$ as a function of $m_\chi$ and $m_H$. To respect these
limits, we set the $He\mu (\tau)$ and $\chi e\mu (\tau)$ Yukawa
couplings to tiny values, order $10^{-9}$, for the range of $m_\chi$
and $m_H$ considered in the present analysis. We are thus left with
$\mu^\pm \tau^\mp $ as the dominant leptonic decay mode of $H(\chi)$.
With the above in mind, we focus on the LFV
signatures of $H$ and $\chi$. To be specific, we focus on two
different types of final state topologies: $\tau_h + 3\mu $ and
$\tau_h +\mu+ 2e $, where $\tau_h$ indicates a hadronically decaying
$\tau$ lepton. We generate the signal events at the leading order for
two representative values of $m_H$ and $m_\chi$ each by varying the
flavor violating Yukawa couplings $Y_\ell$ and $Y_q$ within their
experimentally permissible range.  The corresponding SM background
events are also generated at the leading order. Finally, we obtain the
signal significance using both the cut based and the multivariate
analysis encoded in the boosted decision tree\,(BDT).

The paper is organized in the following way: In Section~\ref{sec:bps},
we outline the choice of benchmark points for the flavour violating
Higgs signal processes. In Section~\ref{sec:signal} and~\ref{sec:bkgs}
we elaborate the signal and various SM background processes. In
Section~\ref{EvSelection}, we present the simulation set up, perform
the collider analysis of the aforementioned two multilepton channels,
and compute the sensitivity of the events at the 14 TeV run of the LHC
experiment. We also compare how the signal cross sections would alter
in the upcoming phase of the 13.6 TeV run of the LHC. We further
demonstrate that for the choices of the Yukawa couplings which are on
the higher side, the signal cross sections and the significance values
corresponding to the already concluded 13 TeV run of the LHC contain
enough incentive for this analysis to be taken up by the experimental
groups for a detailed investigation. In Section \ref{ext},
we examine the impact of including possible systematic uncertainties
on the SM backgrounds for signal benchmarks over an extended range of $m_H$.
Finally we summarize our main findings in Section~\ref{Summary}.

\section{Choice of benchmark points} \label{sec:bps}

First, we focus on various quark and lepton flavor violating couplings
that are involved in the production and decays of $H$ and $\chi$. The
set of new Yukawa couplings that are relevant here are $Y_q$
(i.e. $Y_q^{H,\chi}$) and $Y_\ell$ ($\ell \,= \,\mu~,\tau$), where
$Y_q$ is responsible for the production of $H/\chi$ in $ u ({\bar u})
{\bar c} (c) \to H/\chi$ processes, while $Y_\ell$ drives the decay
$H/\chi \to \mu^\pm \tau^\mp$. Thus the flavor violating signal cross
sections depend upon those two Yukawas $(Y_\ell $ and $Y_q)$ as well
as on $m_\chi$ and $m_H$.

Henceforth, for notational simplicity we shall denote $Y_q^H$ by
$Y_q$.  The values of $Y_q^\chi$ will be automatically set as
$\frac{m_\chi}{m_H} Y_q^H$, as mentioned in the Introduction.  We
produce signal event samples for four different combinations of $(m_H,
m_\chi)$, and for each such combination, we take six benchmark values
for $Y_\ell$ and $Y_q$ each, as shown in Table~\ref{tab:masspoints}.
Those six values for both $Y_\ell$ and $Y_q$ are chosen as
$0.001, 0.003, 0.005, 0.007, 0.009 ~{\rm and} ~0.01$. In view of the \
approximate relation between $Y_q^\chi$ and $Y_q^H \equiv Y_q$, the above six values of quark Yukawa couplings are
consistent with $D^0 - \bar{D}^0$ mixing constraint.

\begin{table}[!h]
\begin{center}
\begin{tabular}{|l|c|c|}
\hline
Mass (GeV) & $m_{\chi} = 20$ & $m_{\chi} = 60$\\
\hline
\texttt{$m_{H} = 160$} & $(Y_{\ell},Y_{q})_{6\times6}$ & $(Y_{\ell},Y_{q})_{6\times6}$  \\
\hline
\texttt{$m_{H} = 170$} & $(Y_{\ell},Y_{q})_{6\times6}$ & $(Y_{\ell},Y_{q})_{6\times6}$ \\
\hline
\end{tabular}
\end{center}
\caption{\small \em Benchmark choices of masses and flavor violating Yukawa couplings.}
\label{tab:masspoints}
\end{table}

Thus there are $4\times 36 = 144$ signal benchmark configurations, and
we calculate the signal significance for each of these sample
points. The values of $m_\chi$ are such that $\chi$ can be produced
on-shell in association with a $Z$ boson from the decay of the heavier
scalar $H$ with sufficient phase space (keeping a mass gap
$\delta m =\, (m_H -m_\chi)\,\sim\,{\cal O}\,(100)$ GeV),
so that the decay products
of $\chi$ and $Z$ have substantial transverse momenta to satisfy the
baseline selection criteria of our analysis. For $m_\chi < 20$ GeV,
the $\chi$ decay products $(\mu, \tau_h)$ would be too soft to be
detected, while for $m_\chi \simeq \left(M_Z, M_W\right)$ the on-shell
two-body decay $H \to \chi + Z$ would be kinematically disfavored for
our choices of $m_H$.

In addition to the above mentioned representative signal benchmark points, we also consider
an extended scenario by varying $m_H$ from $140$ to $500$ GeV, and demonstrate the effects of
systematic uncertainties on signal significance. While doing this analysis, we keep the
range of $m_\chi$ and Yukawa couplings the same as before.

\section{Signal processes} \label{sec:signal}

Having discussed the constraints on our model parameters, we are set
to explore the collider signatures of $H$ and $\chi$ at $\sqrt{s} =
14$ TeV LHC run. We consider two processes {$\bf S1\,(S2)$}, where
$\chi Z (H Z)$ pair is produced on-shell. Subsequently, the
nonvanishing LFV Yukawa coupling $Y_\ell $ induces $\chi(H)$ decay
into $\mu^\pm \tau^\mp$ pairs. From the decay of the $Z$ boson, we
pick up only the $\mu^+\mu^-/e^+ e^-$ final states.
\begin{eqnarray}~\label{Signals}
 {\bf S1:}~p~ p &\to & \chi(\to \mu^\pm \tau^\mp) Z(\to e^+e^-/\mu^+
 \mu^-) \, , \nonumber \\
 {\bf S2:}~p~ p &\to & H (\to \mu^\pm \tau^\mp) Z(\to
 e^+e^-/\mu^+\mu^-) \, . \nonumber 
\end{eqnarray}
\begin{figure}[!h]
\begin{center}
\includegraphics[width= \textwidth]{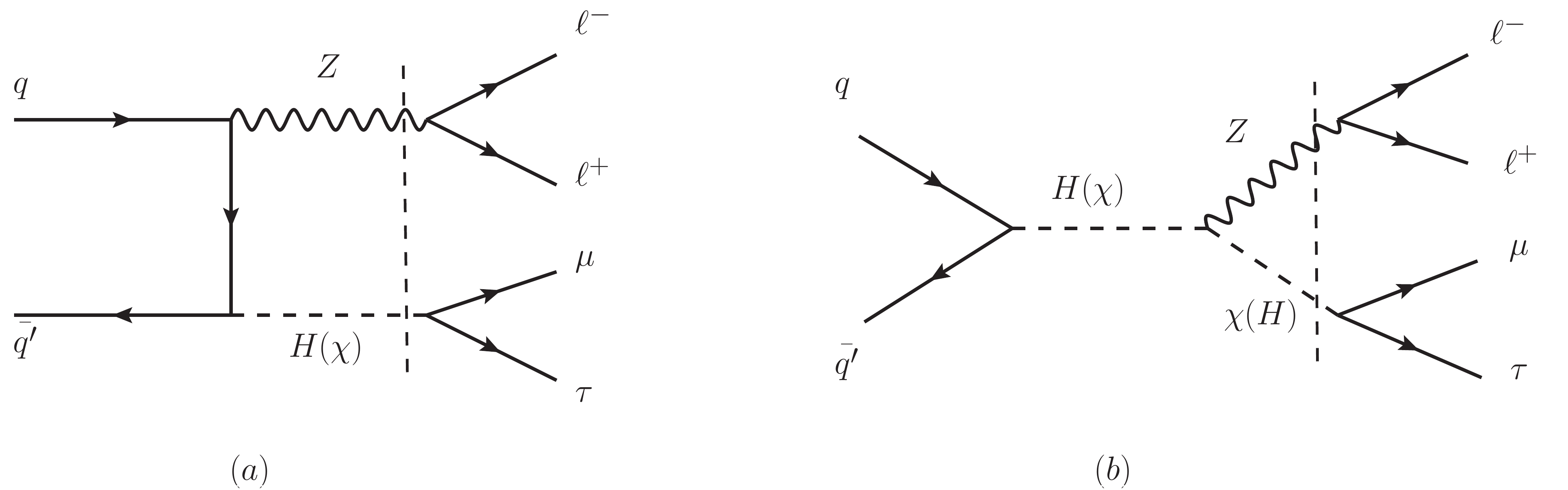}
\caption{\small \em Feynman diagrams for the signal processes. The
  dashed vertical lines are indications of the on-shell production of
  $H/\chi$ in association with $Z$, and their subsequent decays into
  $\mu^{\pm} \tau^{\mp}$ and $\mu^+\mu^-/e^+ e^-$, respectively. }
\label{fig:pp_h2z}
\end{center}
\end{figure}
In the subsequent discussion, whenever we refer to {\bf S1} and/or
{\bf S2} cross section(s), we imply the relevant boson production
cross section(s) $\times$ their branching ratios into multilepton
final satte.  In Figure~\ref{fig:pp_h2z}, we display the {\bf S1} and
{\bf S2} signal processes.  Depending upon the charged lepton flavor
from the $Z$ decay, {\bf S1} and {\bf S2} have the following lepton
flavors in their respective final states: $(a)$ $\tau_h + 3 \mu $ and
$(b)$ $\tau_h + \mu + 2 e$. Treating electron and muon on the same
footing, we are eventually led to $\tau_h + 3 \ell_0 $, $(\ell_0 = e,
\mu) $ as our final signal topology. To simulate the signal we first
implement the LFV Lagrangian in {\tt FeynRules}~\cite{Alloul:2013bka}
to generate a Universal FeynRules Output to be interfaced with the
event generator. We then generate the signal processes using
$\texttt{MadGraph5\_aMC@NLO}$~\cite{Alwall:2014hca} at the leading
order\,(LO). All the SM background events have been
generated using $\texttt{MadGraph5\_aMC@NLO}$. For the calculation of
both the signal and the SM background processes we employ
\texttt{NN23LO1} as the parton distribution
function~\cite{NNPDF:2014otw}. The $\tau$ decays are simulated by the
\texttt{TAUOLA} package integrated in
$\texttt{MadGraph5\_aMC@NLO}$. The parton level events are then passed
through {\tt PYTHIA 8}~\cite{Sjostrand:2014zea} for parton showering,
hadronization and the resulting events are finally processed through
the fast detector simulation package {\tt
  Delphes3}~\cite{deFavereau:2013fsa} using the default CMS card. {\tt
  Delphes} uses the anti-$k_T$ algorithm~\cite{Cacciari:2008gp} to
perform jet clustering using the {\tt FastJet}
package~\cite{Cacciari:2011ma}.  The respective tagging efficiencies
for the $b$ and $\tau$-tagged jets have been parametrically
incorporated within {\tt Delphes}.  We produce signal samples for the
benchmark choices in Table~\ref{tab:masspoints}, which means that 144
signal benchmark points are generated and for each such point we
estimate the signal significance.

\begin{figure}[!h]
\centering
\subfloat[]{
  \label{fig:xsec-20-160}
  \centering
  \includegraphics[width=0.45\textwidth]{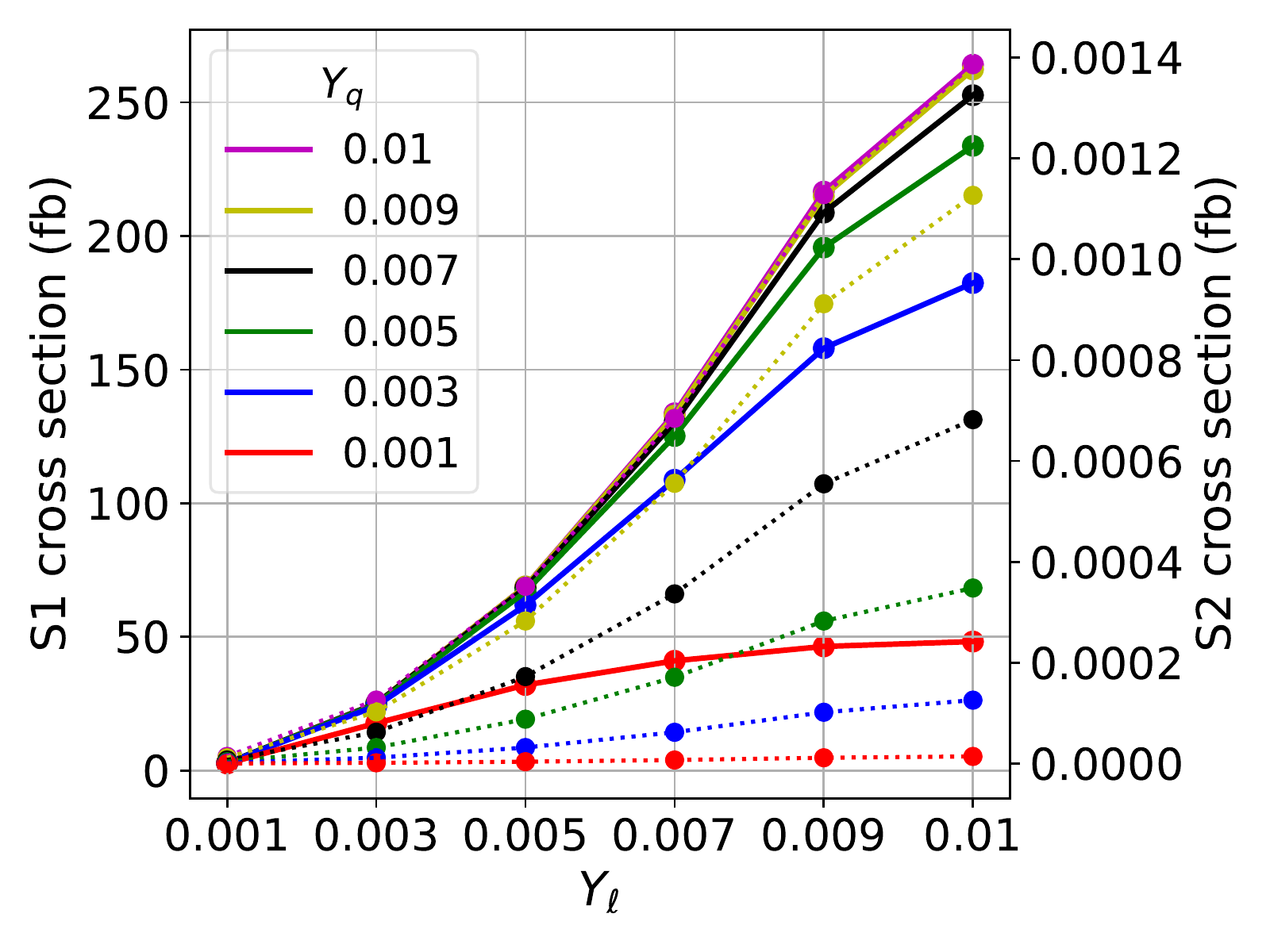}
}
\subfloat[]{
  \label{fig:xsec-60-170}
  \centering
  \includegraphics[width=0.45\textwidth]{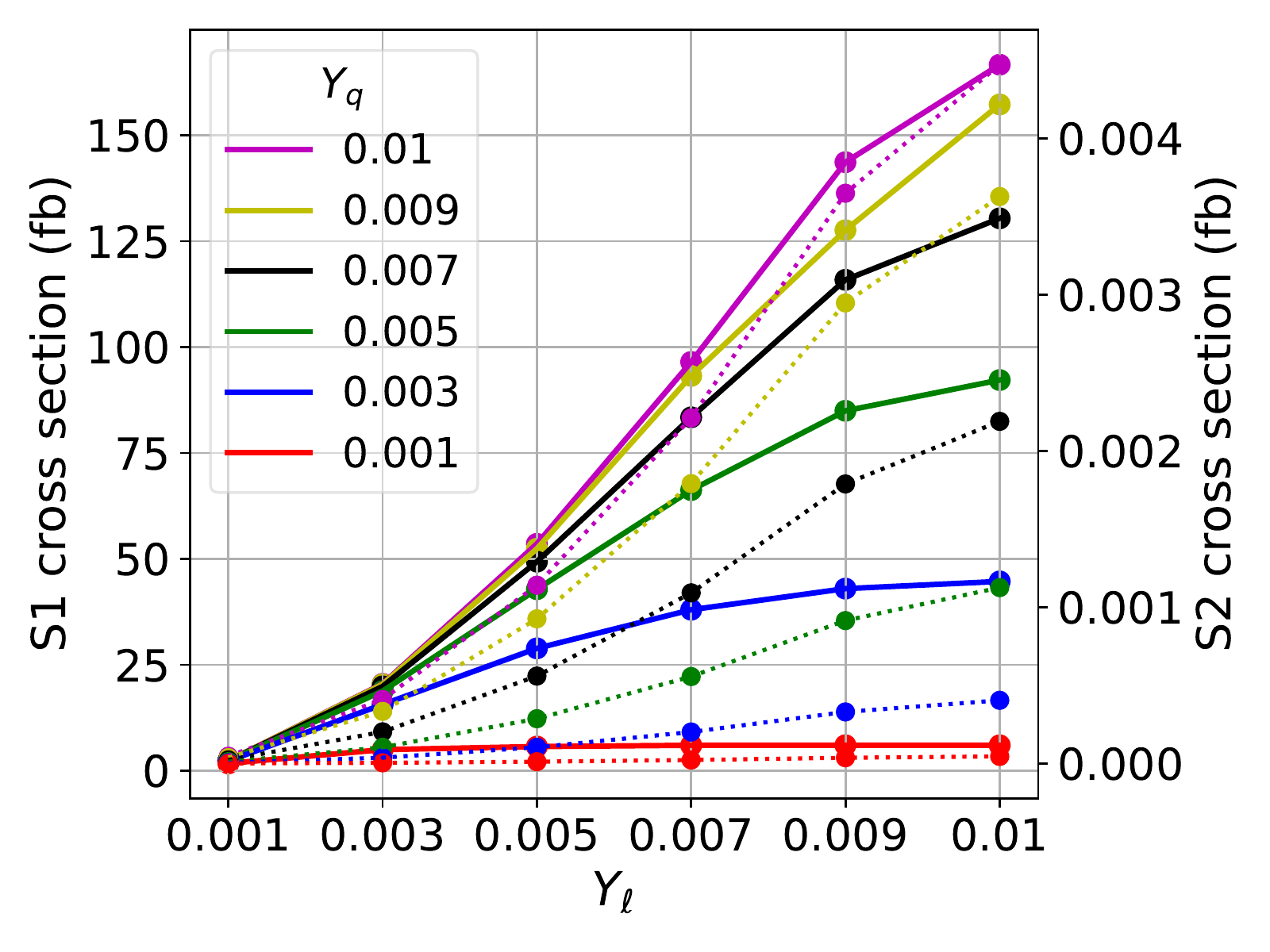}
}
\caption{\small \em Signal process cross sections, i.e. production
  cross sections $\times$ branching ratios at $\sqrt s = 14$ TeV, as a
  function of $Y_{l}$ for fixed values of $Y_{q}$. Solid lines are for
  {\bf S1} and dotted ones are for {\bf S2}. The left panel (a) is for
  $(m_{\chi},m_{H}) \,=\, (20,\,160)$ GeV, while the right panel
  (b) is for $(60,\,170)$ GeV.}
\label{fig:xsec}
\end{figure}

In Figure~\ref{fig:xsec} we exhibit the variation of the cross
sections for {\bf S1} and {\bf S2} as a function of $Y_{\ell}$ for
different values of $Y_{q}$ for the choices, $(m_\chi,m_H)$ in GeV:
(a) $(20,160)$ and (b) $(60,170)$, respectively. We generate all the
signal processes keeping the default \texttt{Madgraph} settings. A few
comments on some gross features of Figure~\ref{fig:xsec} are in order,
as these will have important bearing in planning the collider
analysis.  The {\bf S1} and {\bf S2} cross sections increase with
$Y_q$ and $Y_\ell$. This is expected as $Y_q$ boosts the production
while $Y_\ell$ facilitates the decays of the exotic states into
$\mu^\pm \tau^\mp$. The cross section for $S1$ $(pp \to \chi Z)$ is
roughly five orders of magnitude larger than that of $S2$ $(pp \to H Z)$ simply
because of kinematics $(m_\chi << m_H)$.
Henceforth, we consider only the {\bf S1} process
in our detailed collider analysis. With this signal topology in mind,
we generate appropriate SM backgrounds.

Before leaving this Section, we point out that there exists another
competitive channel worth exploring, which is
\begin{eqnarray}~\label{top_channel}
  p~ p \to t \bar t, t \to b W, \bar t \to u(c) \chi \,(H)\, .
\end{eqnarray}

Here, one of the top quarks decays into $Wb$ via the SM gauge
coupling, while the other can decay into $u(c) \chi $ or $u(c) H$
induced by off-diagonal $H(\chi)tu$ or $H(\chi)tc$ couplings, followed
by $W \to l \nu_l /jj (l = e,\mu,\tau)$ and $\chi \to \mu \tau$. The
cross section for this process is of the same order of magnitude as
that for the associated production channel ({\bf S1}). The {\bf S1}
channel has two distinct advantages over the channel in
Eq.~(\ref{top_channel}) arising from the higher lepton multiplicity in
the final state and from the availability of an invariant mass ($M_Z$)
construction from opposite sign same flavor (OSSF) lepton pairs to be
used as an efficient discriminator against the backgrounds. However,
the process in Eq.~(\ref{top_channel}) merits a separate dedicated
analysis.

A few comments on why we have looked for $\chi$, not singly but in
association with $Z$, are in order. First, from an experimental point
of view, looking for an isolated $\chi$ decaying resonantly to 2-body
($\mu,\,\tau$) final state would be extremely challenging. In spite of
its large cross section, the signal will be completely swamped by the
background. QCD backgrounds for low mass events will be overwhelming,
and also $Z \to \tau^{+}\tau^{-}$ with one $\tau$ decaying to $\mu$
would mimic the signal, specifically for large $m_\chi$, with
uncontrollably large statistics. It is therefore better to focus on
the associated process. Second, from the motivational point of view,
we actually probed a scenario which provides a $H \chi Z$ interaction,
as stressed in the Introduction. Thus a ($\chi Z$) final state,
i.e. production of $\chi$ in association with $Z$ which is easy to
reconstruct with a potential to reign over the background, arises
naturally through the Feynman diagrams shown in Figure
\ref{fig:pp_h2z}.


\section{Backgrounds} \label{sec:bkgs}

There are several SM processes which constitute the background by
imitating the final state $(1\tau^\pm + 3\ell_0)$. For all
backgrounds, we use either the next-to-leading order (NLO) cross
section if they are available in the literature or use the LO cross
sections (weighted by the $k$ factor) employing Madgraph. The dominant
backgrounds arise from $ZZ (\to 4 l^\pm)\,+$ up to 2 jets, $W^\pm Z
(\to 3 l^\pm)\,+$ up to $2$ jets, $t\bar t Z(\to l^+l^-) \,+$ up to 2
jets, $WWZ$ (with all possible decays of $W$ and $Z$), and $t\bar t
\,+$ 1 jet with both top quarks decaying leptonically. The background
from $WZZ$ is subdominant, while negligible contributions arise from
the backgrounds $Z(\to l^+l^-)\,+$ up to $2$ jets, $ZZZ$, $WWW$,
$W^\pm Z \,+$ up to $2$ jets (with $W^\pm \to jj, Z \to l^+ l^- $), $t
\bar t W^\pm \,+$ up to $2$ jets and $ggF/VBF \to h_{125} \to ZZ^* \to
4 l^\pm$.  The LO backgrounds $ZZ + {\rm jets}$, $WZ(3l) + {\rm
  jets}$, $t\bar tZ +{\rm jets}$ are normalised to NLO by the
$k$ factors $1.62$~\cite{Campbell:2011bn}, $1.88$~\cite
{Campbell:2011bn} and $1.4$~\cite{Kardos:2011na}, respectively. For $t
{\bar t} +{\rm jets}$ background, we use the ${\rm N^3LO}$ cross
section from~\cite{Muselli:2015kba}. Rest of the SM background
processes are estimated at LO only.

\section{Analysis} \label{EvSelection}

In this Section we perform a detailed Monte Carlo analysis for the
signal topologies that consist of three charged leptons $(e^\pm
e^\mp\mu^\pm~{\rm or}~ \mu^\pm \mu^\mp \mu^\pm)$ and one tau-tagged
jet $(\tau_{h})$.  We divide the whole event selection procedure into
two steps: $(i)$ Baseline selection and $(ii)$ Signal extraction.

\subsection{Baseline selection} \label{baseline}

Charged particles (leptons and jets) produced in any hard scattering
process at the LHC may not be always visible due to the finite size of
the detector and the requirement of minimum energy to trigger. Hence,
we first apply a set of acceptance cuts (C0) as shown in
Table~\ref{tab:objSel} on all the charged leptons and jets so that
they can be observed at various sub-components of the CMS
detector. Next, we construct various kinematic observables and study
their distributions for both the signal and backgrounds. Based on the
final state composition and distinguishable features of the
distributions of kinematic variables for the signal and backgrounds we
apply preselection cuts (C1$-$C6) to loosely suppress the background
contributions.

\begin{itemize}
\item [] {C0\,:} This consists of basic selection criteria for $e,
  \mu, \tau $ and jets. We use the following set of kinematic
  variables: $(a)$ transverse momentum $p_T$, $(b)$ pseudorapidity
  $\eta$, and $(c)$ angular separation between two objects, $\Delta
  R$, where $\Delta R_{ij} = \sqrt{(\Delta \eta)^2 + (\Delta \Phi)^2}$
  is defined in terms of the azimuthal angular separation $(\Delta
  \Phi)$ and pseudorapidity difference $(\Delta \eta)$ between two
  objects $i$ and $j$. The threshold values of these variables are
  shown in Table~\ref{tab:objSel}.

\begin{table}[!h]
\begin{center}
\begin{tabular}{|l|l|}
  \hline Objects & Selection cuts \\
  \hline \texttt{$e$} & $p_{T} > 10$~{\rm GeV}, $~|\eta| < 2.5$, $~\Delta R_{e \mu} > 0.4$\\
  \texttt{$\mu$} & $p_{T} > 10$~{\rm GeV}, $~|\eta| < 2.4$, $~\Delta R_{e \mu} > 0.4$ \\
  \texttt{$\tau_{h}$} & $p_{T} > 20$~{\rm GeV}, $~|\eta| < 2.4$, $~\Delta R_{\tau, e\mu} > 0.4$ \\
  \texttt{Jet} & $p_{T} > 20$~{\rm GeV}, $~|\eta| < 4.7$, $~\Delta R_{{\rm jet}, e\mu} > 0.4$ \\
  \hline
\end{tabular}
\end{center}
\caption{\small \em Summary of acceptance cuts.}
\label{tab:objSel}
\end{table}

\item [] {C1\,:} The signal has a $Z$ boson decaying to a pair of OSSF
  leptons $(e,\mu)$. To ensure the presence of one $Z$ boson, we
  select events with an invariant mass $M_{\ell_0^+ \ell_0^-}$ close
  to the $Z$ peak by demanding $\mid M_{\ell_0^+\ell_0^-}-M_Z\mid\, <
  10$ GeV, where $M_Z$ is the true $Z$ mass. The same cut has been
  used to suppress the SM di-$Z$ contribution by rejecting events
  having more than one $Z$ boson.

 \item [] {C2\,:} We look only for the leptonic decay of $\chi$,
   i.e. $\chi \to \mu^{\pm} +\tau^{\mp}$. So, we require at least one
   $\mu$ to be present in the selected events.

 \item [] {C3\,:} Based on C1 and C2, we demand the presence of three
   charged leptons in the final state, one $\mu$ from $\chi$ decay and
   $e^+ e^- / \mu^+\mu^-$ from $Z$ decay.

 \item [] {C4\,:} One of the decay products of $\chi$ is a $\tau$
   lepton, and we choose to work with the $\tau$ that decays in
   hadronic mode. We require at least one $\tau$ jet ($\tau_{h}$) to
   be present in the final state.
 
 \item [] {C5\,:} Now we have three muons (with one OSSF pair), or,
   one muon $+$ one OSSF electron pair in the final state, along with
   the $\tau$ jets. The $\mu$ which is not a decay product of $Z$,
   paired with a $\tau_h$ of opposite charge. Together they are
   perceived to have arisen from $\chi$ decay. This $\mu$ is denoted
   as $\mu^\prime$ hereafter.
 
 \item [] {C6\,:} The signal final state is free from $b$ jets. So, we
   impose a $b$ jet veto in our baseline selection to suppress the top
   quark enriched SM backgrounds.
\end{itemize}

\begin{table}[h!]
\begin{center}
{\footnotesize
    \centering
    \setlength{\tabcolsep}{0.5em} 
{\renewcommand{\arraystretch}{1.3}
    \begin{tabular}{|>{}c|*{9}{c|}}\hline
      \multirow{2}{*}{Samples} & \multirow{2}{*}{$\sigma_{\rm prod}$ $\times$} & \multicolumn{7}{c|}{Effective cross sections (fb)} & \multirow{2}{*}{Events} \\\cline{3-9}
      &BR (fb) &C0 &C1 &C2 & C3 & C4 & C5 & C6 &($300\,{\rm fb^{-1}}$) \\ \hline
      \texttt{Signal } & & & & & & & & & \\
      \texttt{$(m_\chi, m_H)-(Y_\ell, Y_q)$} & & & & & & & & & \\
      \texttt{$(20,160)-(0.003,0.001)$} & $2.8$ & $2.708$ & $0.747$ & $0.620$ & $0.367$ & $0.037$ & $0.036$ & $0.035$ & $10.47$ \\
      \texttt{$(20,160)-(0.005,0.005)$} & $66.89$ & $64.799$ & $17.894$ & $14.865$ & $8.809$ & $0.895$ & $0.874$ & $0.845$ & $253.53$ \\
      \texttt{$(20,160)-(0.009,0.007)$} & $132.82$ & $128.59$ & $35.472$ & $29.428$ & $17.403$ & $1.818$ & $1.772$ & $1.725$ & $517.6$ \\
      \hline
      \texttt{$(20,170)-(0.003,0.001)$} & $2.3$ & $2.238$ & $0.627$ & $0.522$ & $0.310$ & $0.037$ & $0.036$ & $0.035$ & $10.54$ \\
      \texttt{$(20,170)-(0.005,0.005)$} & $55.44$ & $53.945$ & $15.278$ & $12.748$ & $7.493$ & $0.89$ & $0.878$ & $0.845$ & $253.47$ \\
      \texttt{$(20,170)-(0.009,0.007)$} & $110.16$ & $107.19$ & $29.999$ & $25.015$ & $14.718$ & $1.772$ & $1.737$ & $1.667$ & $500.18$ \\
      \hline
      \texttt{$(60,160)-(0.003,0.001)$} & $2.98$ & $2.89$ & $0.701$ & $0.637$ & $0.464$ & $0.079$ & $0.077$ & $0.074$ & $22.3$ \\
      \texttt{$(60,160)-(0.005,0.005)$} & $53.55$ & $51.914$ & $12.508$ & $11.361$ & $8.297$ & $1.38$ & $1.353$ & $1.304$ & $391.08$ \\
      \texttt{$(60,160)-(0.009,0.007)$} & $114.44$ & $110.88$ & $26.875$ & $24.375$ & $17.768$ & $2.981$ & $2.921$ & $2.82$ & $845.94$ \\
      \hline
      \texttt{$(60,170)-(0.003,0.001)$} & $2.29$ & $2.231$ & $0.58$ & $0.523$ & $0.368$ & $0.066$ & $0.065$ & $0.063$ & $18.75$ \\
      \texttt{$(60,170)-(0.005,0.005)$} & $42.83$ & $41.717$ & $10.802$ & $9.737$ & $6.887$ & $1.203$ & $1.181$ & $1.142$ & $342.55$ \\
      \texttt{$(60,170)-(0.009,0.007)$} & $93.2$ & $90.751$ & $23.49$ & $21.222$ & $14.988$ & $2.669$ & $2.619$ & $2.521$ & $756.18$ \\
      \hline\hline\hline
      \texttt{SM backgrounds} & & & & & & & & & \\
      \texttt{$Z + {\rm jets}$} & $6.33\times10^{6}$ & $6.32\times10^{6}$ & $2.9\times10^{5}$ & $1.8\times10^{5}$ & $12.74$ & $0.11$ & $0$ & $0$ & $0$ \\
      \hline
      \texttt{$t \bar t + {\rm jets ~(2l)}$} & $1.09\times10^{5}$ & $1.09\times10^{5}$ & $1522.34$ & $967.5$ & $3.58$ & $0.1$ & $0.03$ & $0.03$ & $9.11$ \\
      \hline
      \texttt{$t \bar t W^{\pm} + {\rm jets}$} & $253.8$ & $253.78$ & $1.125$ & $0.779$ & $0.22$ & $0.013$ & $0.005$ & $0.001$ & $0.43$ \\
      \texttt{$t \bar t Z + {\rm jets}$} & $240.3$ & $240.3$ & $57.68$ & $39.79$ & $11.86$ & $1.193$ & $0.536$ & $0.141$ & $42.15$\\
      \hline
      \texttt{$W^{\pm}Z + {\rm jets ~(3l)}$} & $2273$ & $2263.6$ & $849.86$ & $614.95$ & $389.99$ & $3.67$ & $1.207$ & $1.144$ & $343.17$ \\
      \texttt{$W^{\pm}Z + {\rm jets ~(2l)}$} & $4504$ & $4496.3$ & $1220.17$ & $769.65$ & $0.18$ & $0.007$ & $0.002$ & $0.002$ & $0.55$ \\
      \texttt{$ZZ + {\rm jets ~(4l)}$} & $187.3$ & $186.46$ & $71.86$ & $51.86$ & $26.89$ & $2.106$ & $1.286$ & $1.254$ & $376.34$ \\
      \texttt{${\rm [GGF]}~ ZZ ~{\rm (4l)}$} & $14.82$ & $14.476$ & $2.16$ & $1.68$ & $0.92$ & $0.01$ & $0.003$ & $0.002$ & $0.73$ \\
      \texttt{${\rm [VBF]}~ ZZ ~{\rm (4l)}$} & $2.211$ & $2.21$ & $0.32$ & $0.24$ & $0.13$ & $0.003$ & $0.001$ & $0.001$ & $0.28$ \\
      \hline
      \texttt{$WWW$} & $236.2$ & $236.07$ & $0.6$ & $0.39$ & $0.08$ & $0$ & $0$ & $0$ & $0$ \\
      \texttt{$WWZ$} & $188.9$ & $188.75$ & $4.84$ & $3.24$ & $1.0$ & $0.07$ & $0.038$ & $0.034$ & $10.2$ \\
      \texttt{$WZZ$} & $63.76$ & $63.65$ & $3.036$ & $2$ & $0.46$ & $0.025$ & $0.01$ & $0.009$ & $2.64$ \\
      \texttt{$ZZZ$} & $15.8$ & $15.73$ & $1.08$ & $0.69$ & $0.06$ & $0.007$ & $0.004$ & $0.003$ & $0.95$ \\
      \hline
    \end{tabular}}}
    \end{center}
    \caption{\small \em The signal and SM background effective cross
      sections (fb) after each successive baseline cut (C0-C6) and
      final event yields for ${\cal L} = 300~ fb^{-1}$ at 14 TeV
      LHC run. Signal event samples are generated for a few
      representative values of $m_{\chi}$ and $m_{H}$ (in GeV) and for
      a range of $Y_{\ell}$ and $Y_{q}$. Signal cross sections are
      calculated at LO, $t \bar t +$ jets cross sections at N$^{3}$LO,
      while the other SM backgrounds are estimated at NLO.}
    \label{tab:baselineYield}
\end{table}

\begin{figure}[!h]
\centering
\subfloat[]{
  \label{fig:xmupt}
	\centering
  \includegraphics[width=0.45\textwidth]{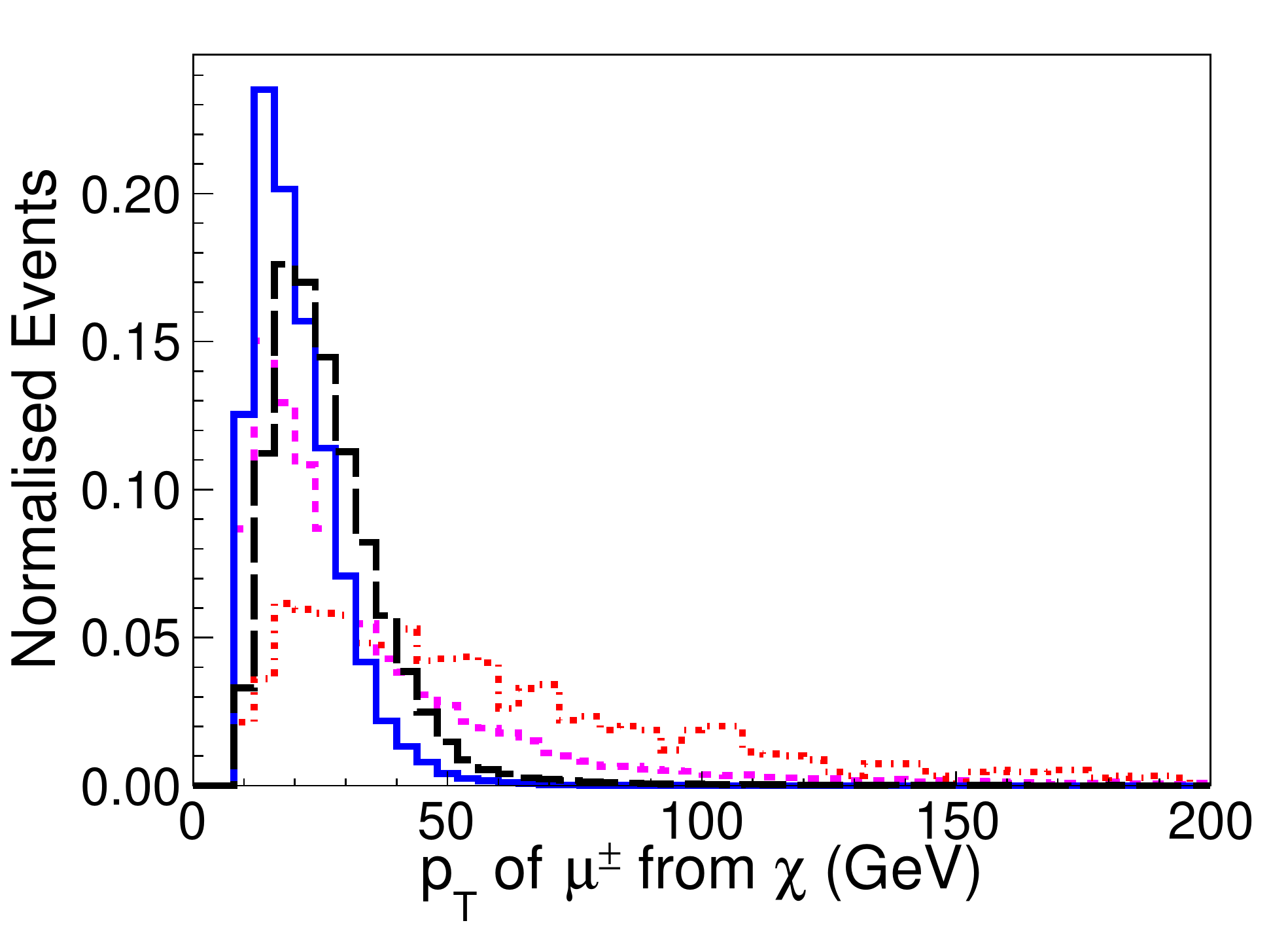}
}
\subfloat[]{
  \label{fig:xtaupt}
	\centering
  \includegraphics[width=0.45\textwidth]{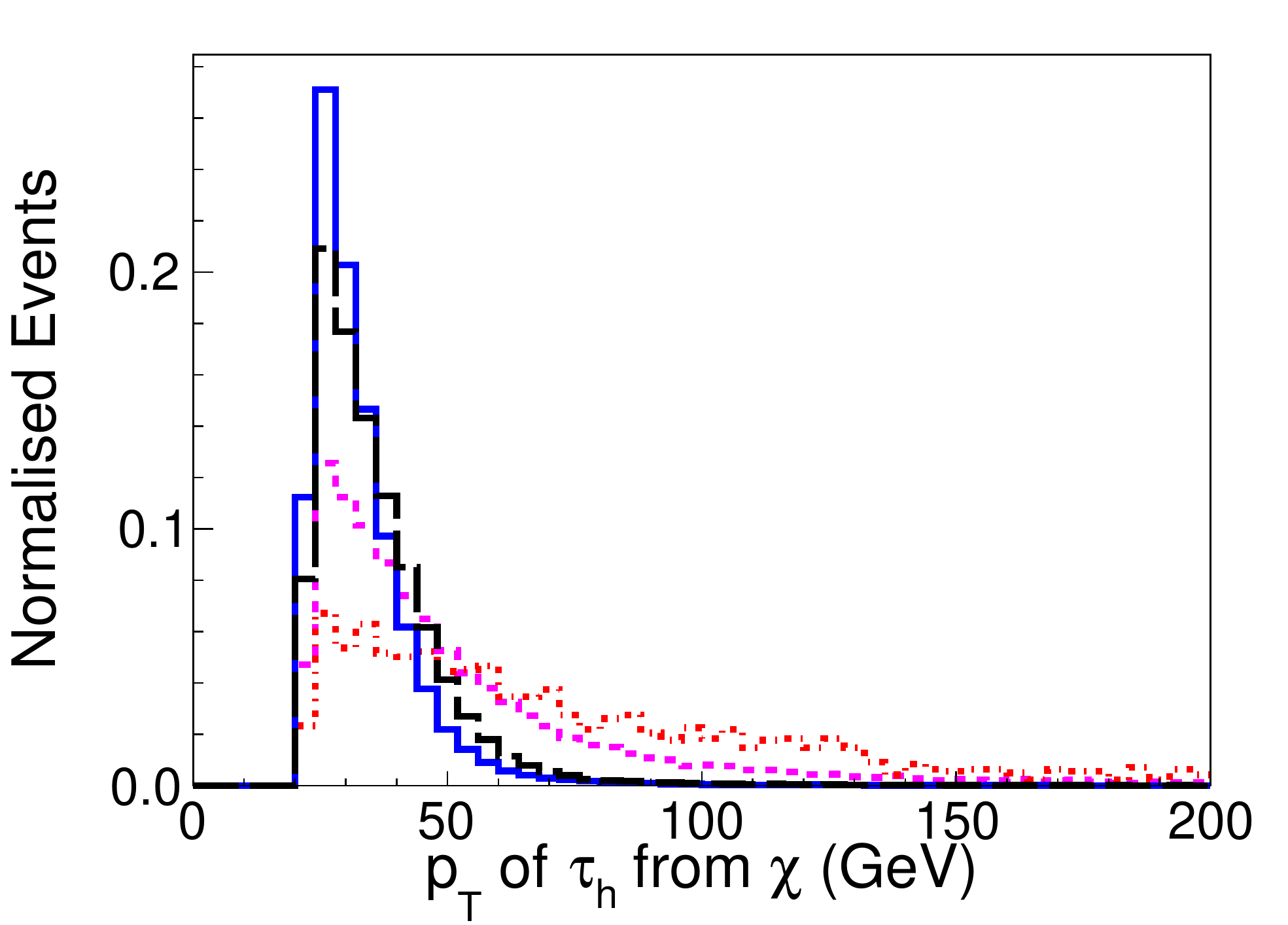}
}
\hspace{0.01\textwidth}
\subfloat[]{
  \label{fig:met}
	\centering
  \includegraphics[width=0.45\textwidth]{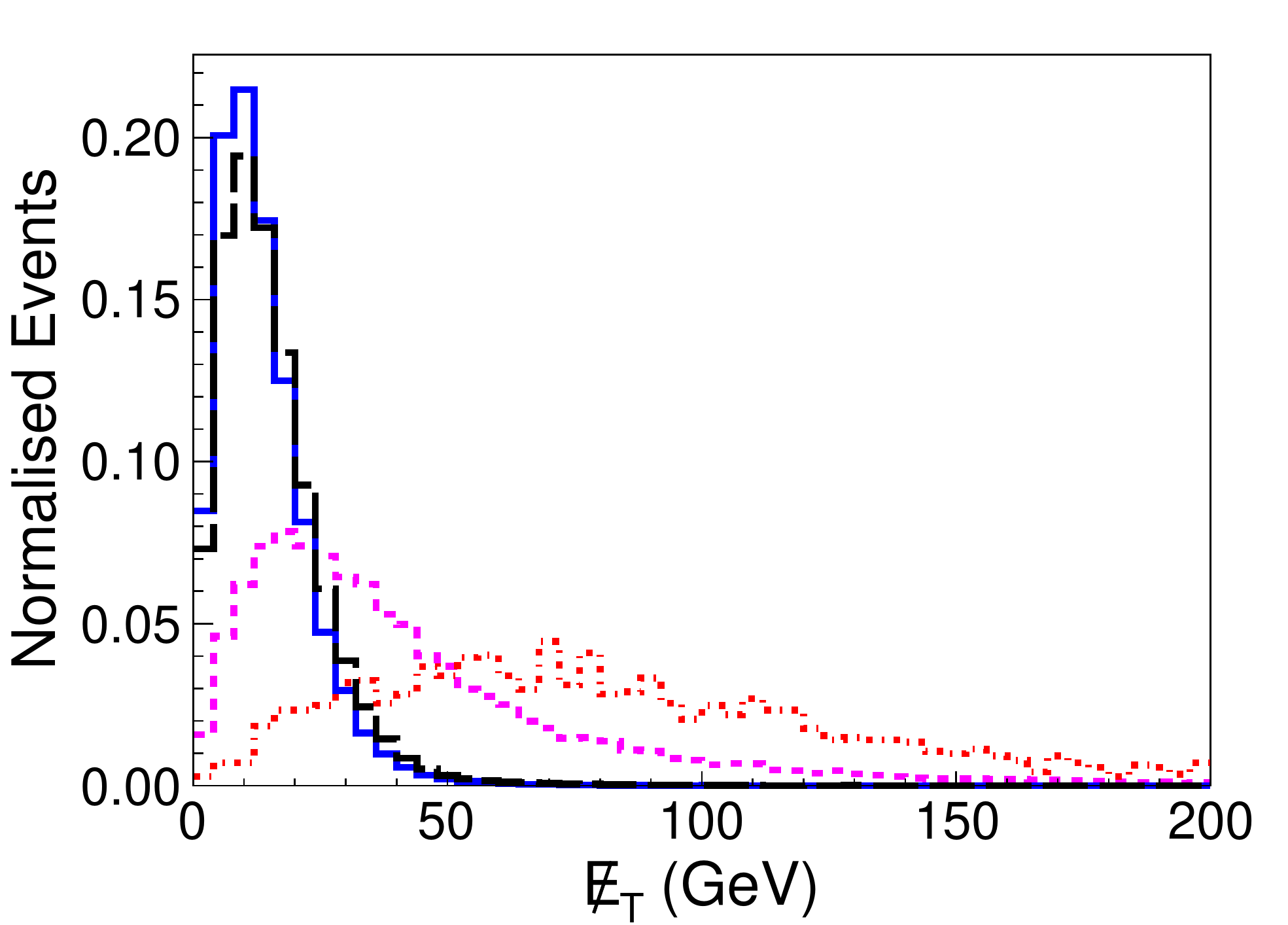}
}
\subfloat[]{
  \label{fig:drzl1l2}
	\centering
  \includegraphics[width=0.45\textwidth]{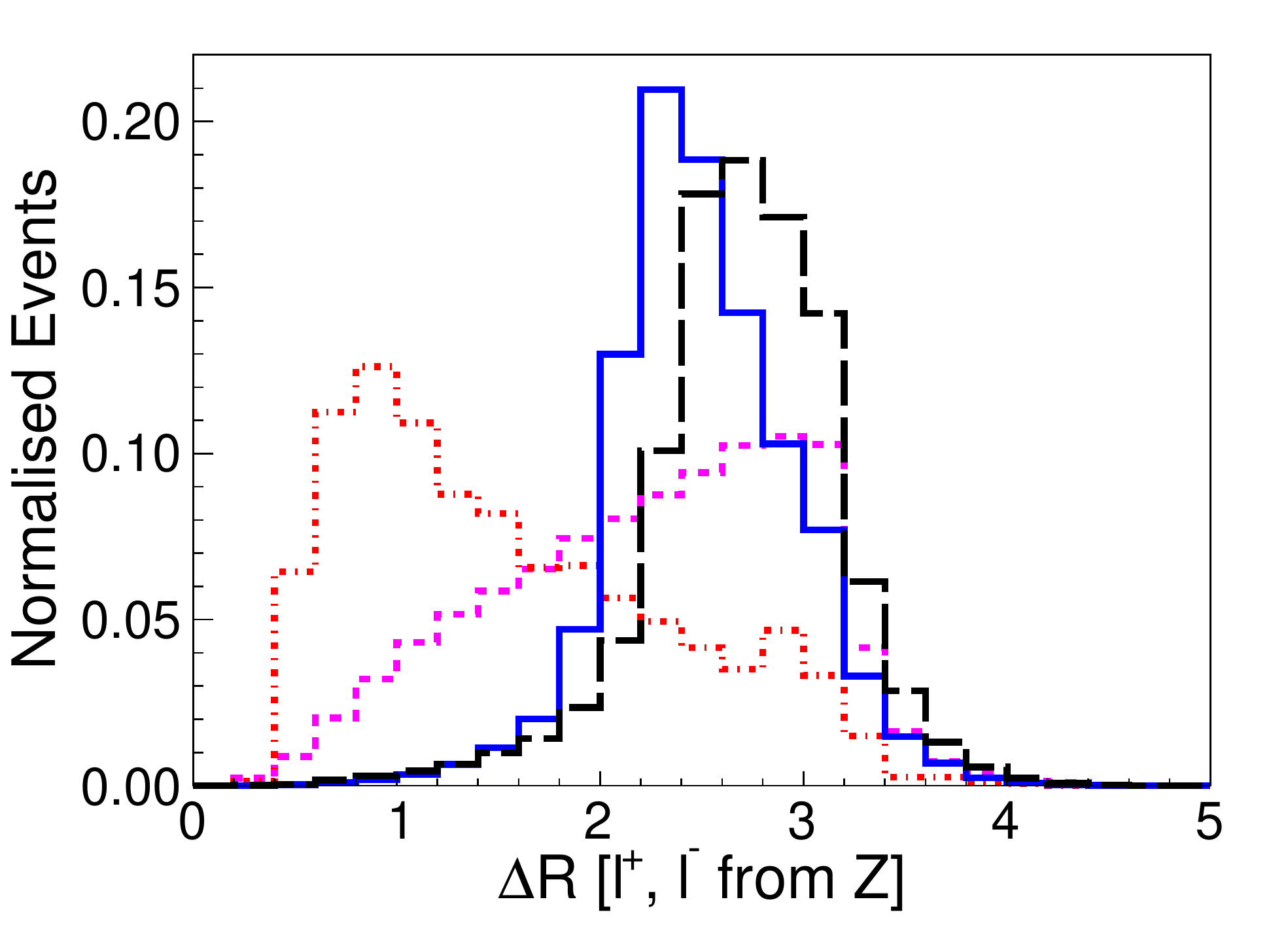}
}
  \label{fig:legend}
	\centering
  \includegraphics[width=0.75\textwidth]{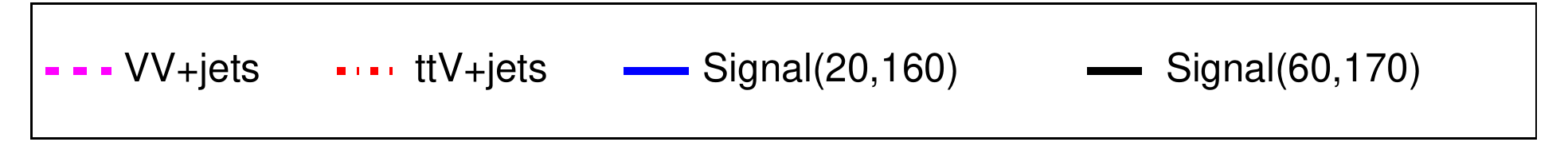}
\caption{\small \em Distributions of some kinematic variables, (a)
  $p_T$ of $\mu$ (i.e. $\mu^\prime$) from $\chi$, (b) $p_T$ of
  $\tau_h$ from $\chi$, (c) $\met$ and (d) $\Delta R$ between the two OSSF leptons from
  $Z$, after baseline selection are shown for two representative mass points $(m_{\chi},m_H)$ in
  GeV: $(20,160)$ [solid blue] and $(60, 170)$ [black long dashed]. For
  comparison, distributions of the same kinematic variables are shown
  for two major SM backgrounds, namely, $VV+~{\rm jets}$ [magenta
    small dashed] and $ttV+~{\rm jets}$ [red dash-dotted], where $V =
  W^\pm, Z$.}
\label{fig:features}
\end{figure}

In Table~\ref{tab:baselineYield} we present the effective cross
section (fb) after acceptance and successive pre-selection cuts, C1 to
C6, for both the signal and background events, and in the last column
we show the corresponding number of events at an integrated luminosity
of $300\,{\rm fb}^{-1}$. One should note that we only show a few
representative benchmark points for the signal samples. One can see
from Table~\ref{tab:baselineYield} that after the baseline selection
cuts, major SM background processes turn out to be $ZZ\,(4 l) +\,{\rm
  jets}$ and $W^{\pm} Z\,(3l) +\,{\rm jets}$, followed by $t\bar t Z +
{\rm jets}$. The $ZZ ~(4l) + ~{\rm jets}$ process has two $Z$ bosons
decaying to leptons of any flavor.  The hadronic branching ratio of
$\tau$ is $64.8\%$~\cite{ParticleDataGroup:2018ovx} and the detection
efficiency of such a $\tau$ is $60\%$ as considered in
$\texttt{Delphes}$. The combined effect of these two is the main
reason for getting low signal efficiency of C4 cut. In
Figure~\ref{fig:features} we show normalized distributions of a few
kinematic variables for the signal and two most dominant classes of SM
backgrounds after the baseline selection. For signal events, we choose
two benchmark mass points $(m_\chi, m_H) = (20,160)$ and
$(60,170)~{\rm GeV}$. The dominant background processes are $VV+~{\rm
  jets}$ ($VV = WZ, ZZ$) and $ttV+~{\rm jets}$ ($V = W, Z$).
These variables shown in Figure~\ref{fig:features} would play a significant r\^ole
in the signal discrimination both in the cut based and the multivariate analyses.
The major source of \,$\met$ in the signal is
from neutrino produced in the hadronic $\tau$ decay.  One should note
that undetected charged leptons and/or $\tau$ jet are likely to make
small contributions to the missing transverse energy. Since the
choices for the mass $m_\chi$ of the parent particle in the two cases
are not very different, we do not expect any significant change in the
$\met$ distributions for two different signal benchmark points, as
depicted in Figure~\ref{fig:met}.  The transverse momentum ($p_T$) of
$\mu^\prime$ and $\tau_h$ show similar behavior as $\met$, as shown in
Figures~\ref{fig:xmupt} and~\ref{fig:xtaupt}, respectively. All the
three distributions show that the SM backgrounds are harder than the
signal contributions. We also show the
distribution of $\Delta R$ between the two OSSF leptons i.e. the
decay products of $Z$ in Figure~\ref{fig:drzl1l2}.

At the end of the baseline selection, we attempt to reconstruct
$m_\chi$ by combining the four momenta of its decay products,
$\mu^\prime $ and $\tau_h$.  Unfortunately, this prescription does not
work in this case because one cannot fully reconstruct the four
momentum of $\tau$ as the hadronic decay of tau is associated with a
missing neutrino. Nevertheless, one can still get some idea about
$m_\chi$ using a different kinematic variable, transverse mass
$(m_T)$, defined in terms of $p_T$ of $\mu^\prime, \tau_h$ and
$\met$. In doing so, we also assume that the aforementioned neutrino
is the only source of $\met$ for signal events. For a two body decay,
the transverse mass is defined as:
 \begin{equation}~\label{eq:general_mT}
   m_{T} (a,b) = \sqrt{2 \times p_{T}^{a} \times p_{T}^{b} \times (1 -
     \cos (\Delta\Phi^{a,b}))} \, ,
 \end{equation}
\noindent
where $a$ and $b$ are the final decay products and $\Delta\Phi^{a,b}
=\,\mid \Phi_a - \Phi_b\mid $. For the final state considered here,
the transverse mass variable for the system comprising $\mu^{\prime}$,
$\tau_{h}$ and $\met$ is constructed
as~\cite{ATLAS:2014vhc,CMS:2018rmh}:
 \begin{equation}~\label{eq:comp_mT}
   m_{T} (\mu^{\prime}, \tau_{h}, \met) = \sqrt{m_{T}^{2}
     (\mu^{\prime}, \met) + m_{T}^{2} (\tau_{h},
     \met) + m_{T}^{2} (\mu^{\prime}, \tau_{h})} \, .
 \end{equation}

\begin{figure}[!h]
\centering
\subfloat[]{
  \label{fig:mT}
	\centering
  \includegraphics[width=0.45\textwidth]{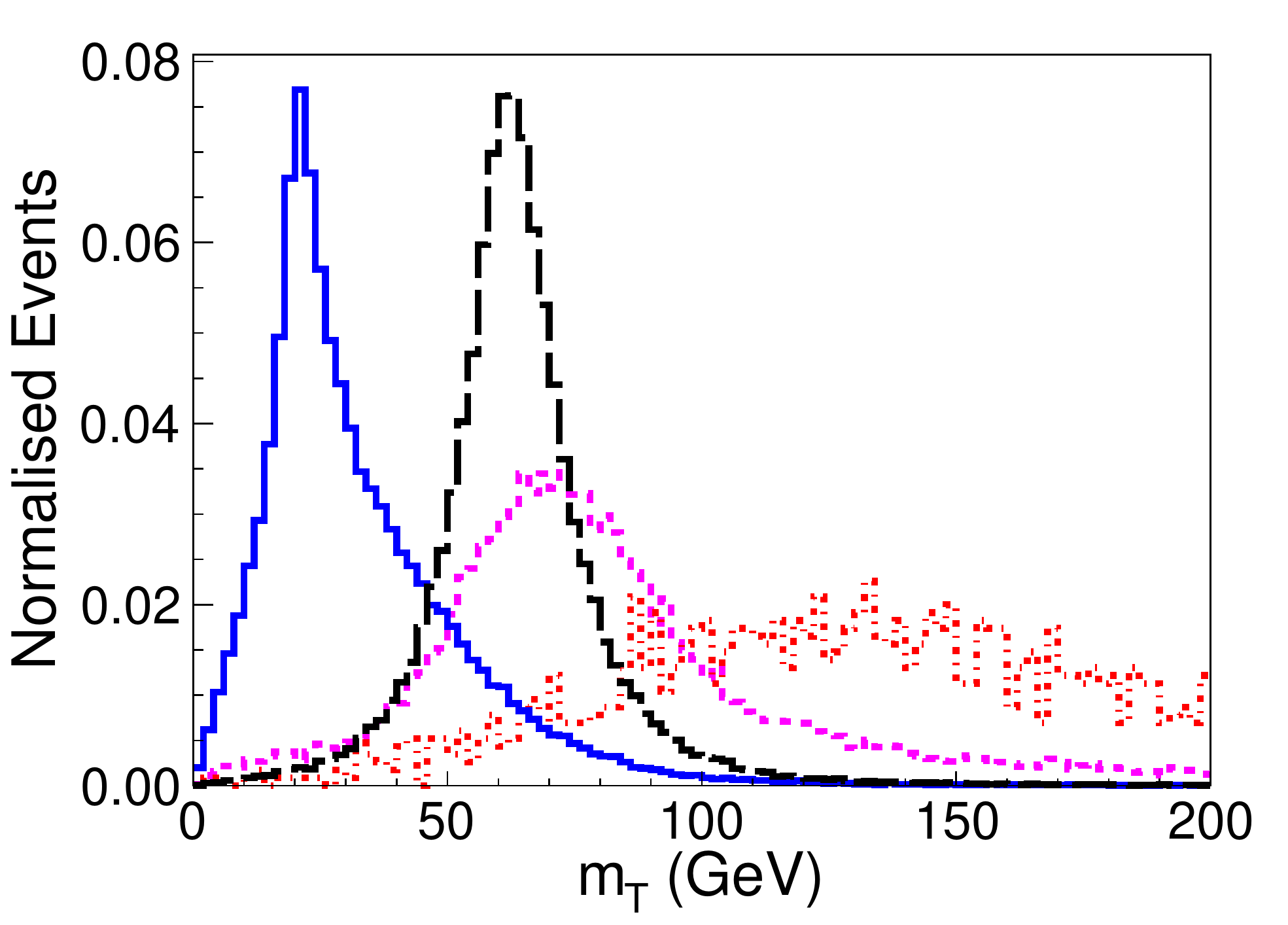}
}
\subfloat[]{
  \label{fig:mcol}
	\centering
  \includegraphics[width=0.45\textwidth]{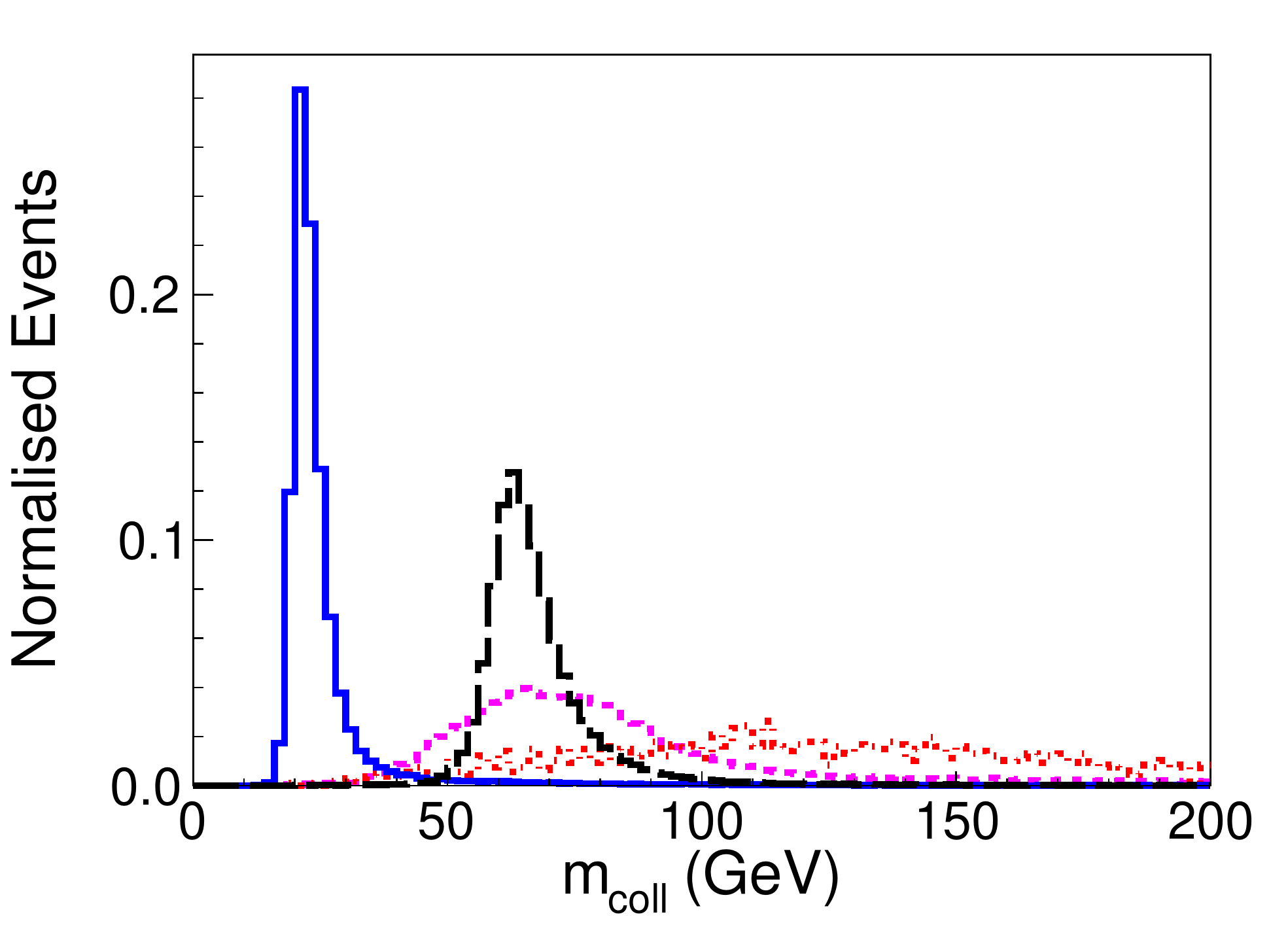}
}
  \label{fig:legend1}
	\centering
  \includegraphics[width=0.7\textwidth]{Legend.pdf}
\caption{\small \em Distributions of (a) transverse mass ($m_{T}$) and
  (b) collinear mass ($m_{coll}$) of the reconstructed $\chi$ after
  baseline selection.}
\label{fig:mTs}
\end{figure}

In Figure~\ref{fig:mT} we show normalized transverse mass
distributions for both the signal and the SM backgrounds. Here we
clearly see the presence of reconstructed $\chi$ around 20 GeV and 60
GeV. The peak of the signal distribution is very well separated from
the SM backgrounds $VV+$ jets as long as $m_\chi$ is not close to
either $M_W $ or $M_Z$. We observe that for $m_\chi = 60$ GeV, there
is a substantial overlap between the signal region and the SM
backgrounds $VV+$ jets.  Figure~\ref{fig:mcol} shows the distribution
of collinear mass~\cite{Ellis:1987xu} of the $\mu^\prime$, $\tau_h$
and $\met$ system, which we explain in the following. The collinear
mass technique is useful to reconstruct the mass of a particle
decaying to $\tau$ and other visible objects. It is assumed that the
decay products of $\tau$ are boosted in the original direction of
$\tau$ itself since $m_\tau \ll m_\chi$. Thus the transverse component
of the $\tau$ neutrino ($\nu_\tau$) momentum, $p_{T}^{\nu}$, can be
estimated by taking the projection of $\met$ in the direction of
visible $\tau_h$.  The definition of the collinear mass is
\begin{equation}~\label{colmass}
    m_{coll} = \frac{m_{vis}}{\sqrt \beta_{\tau}^{vis}} \, , 
\end{equation}
\noindent
where $m_{vis}$ represents the invariant mass of $\tau_h$ and
$\mu^\prime$, whereas $\beta_{\tau}^{vis}$ is the fraction of the
$\tau$ momentum carried by the visible (hadronic) $\tau$ decay
products ($\tau_h$) i.e. $\frac{p_{T}^{\tau_h}}{p_{T}^{\tau_h} +
  p_{T}^{\nu}}$~\cite{CMS:2021rsq}.  The collinear mass exhibits a
somewhat better resolution than the transverse mass distribution. Both
distributions affirm the existence of $\chi$ in the signal processes.
We have performed the exercise to indicate the possibility as well as
the limitations of the mass reconstruction procedure in our scenario.
We further extend and refine our baseline analysis to cut based and
multivariate analyses to obtain the final signal significance ${\cal
  S}$ defined in terms of number of signal and background events $S$
and $B$ as:
\begin{equation}~\label{sigEq}
    {\cal S} = \frac{S}{\sqrt{S + B}} \, , 
\end{equation}
\noindent
where $S(B)$ can be estimated as: $S(B) = \sigma_{S(B)} \times {\cal
  L} \times \epsilon_{S(B)}$, with $\sigma_{S(B)}$, ${\cal L}$ and
$\epsilon_{S(B)}$ denoting the signal (background) cross section,
integrated luminosity and signal (background) selection efficiency,
respectively.

\subsection{Signal extraction : Cut-based analysis} \label{cutbasedanalysis}

In the cut based analysis, we impose a condition of
missing energy on top of the baseline selection. As the  background can have
several sources of $\met$, namely, multiple neutrinos, jet energy
mismeasurements and mistagged charged leptons, it has a much harder
$\met$ spectrum compared to the signal. We reject all events with
$\met > 40 \,{\rm GeV}$ and this substantially reduces various
background contributions.

\begin{table}[!h]
  \begin{center}
    {\footnotesize
      \centering
      \setlength{\tabcolsep}{0.5em}
                {\renewcommand{\arraystretch}{1.3}
      \begin{tabular}{|>{}c|*{4}{c|}}
      \hline
      \multirow{2}{*}{Samples} & \multirow{2}{*}{Baseline}       & \multirow{2}{*}{Effective cross sections (fb)} & \multirow{2}{*}{Events} \\
                               &    cross section (fb)           &             $(\met\,<\,40\,{\rm GeV})$         & ${\cal L}=300\, {\rm fb}^{-1}$ \\
        \hline
        \texttt{Signals} & & & \\
        \texttt{$(m_\chi, m_H)-(Y_\ell, Y_q)$} & & & \\
        \texttt{$(20,160)-(0.003,0.001)$} & $0.035$ & $0.034$ & $10.2$    \\
        \texttt{$(20,160)-(0.005,0.005)$} & $0.845$ & $0.832$ & $249.6$   \\
        \texttt{$(20,160)-(0.009,0.007)$} & $1.725$ & $1.697$ & $509.1$   \\
        \hline
        \texttt{$(20,170)-(0.003,0.001)$} & $0.035$ & $0.034$ & $10.2$    \\
        \texttt{$(20,170)-(0.005,0.005)$} & $0.845$ & $0.826$ & $247.8$   \\
        \texttt{$(20,170)-(0.009,0.007)$} & $1.667$ & $1.64$  & $491.9$   \\    
        \hline\hline
        \texttt{$(60,160)-(0.003,0.001)$} & $0.074$ & $0.073$ & $21.9$   \\
        \texttt{$(60,160)-(0.005,0.005)$} & $1.304$ & $1.278$ & $383.4$  \\
        \texttt{$(60,160)-(0.009,0.007)$} & $2.82$  & $2.762$ & $828.6$  \\      
        \hline
        \texttt{$(60,170)-(0.003,0.001)$} & $0.063$ & $0.061$ & $18.3$   \\
        \texttt{$(60,170)-(0.005,0.005)$} & $1.142$ & $1.114$ & $334.2$  \\
        \texttt{$(60,170)-(0.009,0.007)$} & $2.521$ & $2.445$ & $733.5$  \\
        \hline \hline
        \texttt{SM backgrounds} & & & \\
        \texttt{$t \bar t  Z + {\rm jets}$}    & $0.141$ & $0.025$ & $7.5$    \\
        \texttt{$W^{\pm}Z + {\rm jets (3l )}$} & $1.144$ & $0.423$ & $126.9$  \\
        \texttt{$ZZ + {\rm jets ~(4l )}$}      & $1.254$ & $0.816$ & $244.8$  \\
        \hline
    \end{tabular}}}
  \end{center}
  \caption{\em Summary of the cut based analysis.}
  \label{tab:cutYield}
\end{table}

\begin{figure}[!tbh]
\centering
\subfloat[]{
  \label{fig:cut-sign-20-160}
	\centering
  \includegraphics[width=0.49\textwidth]{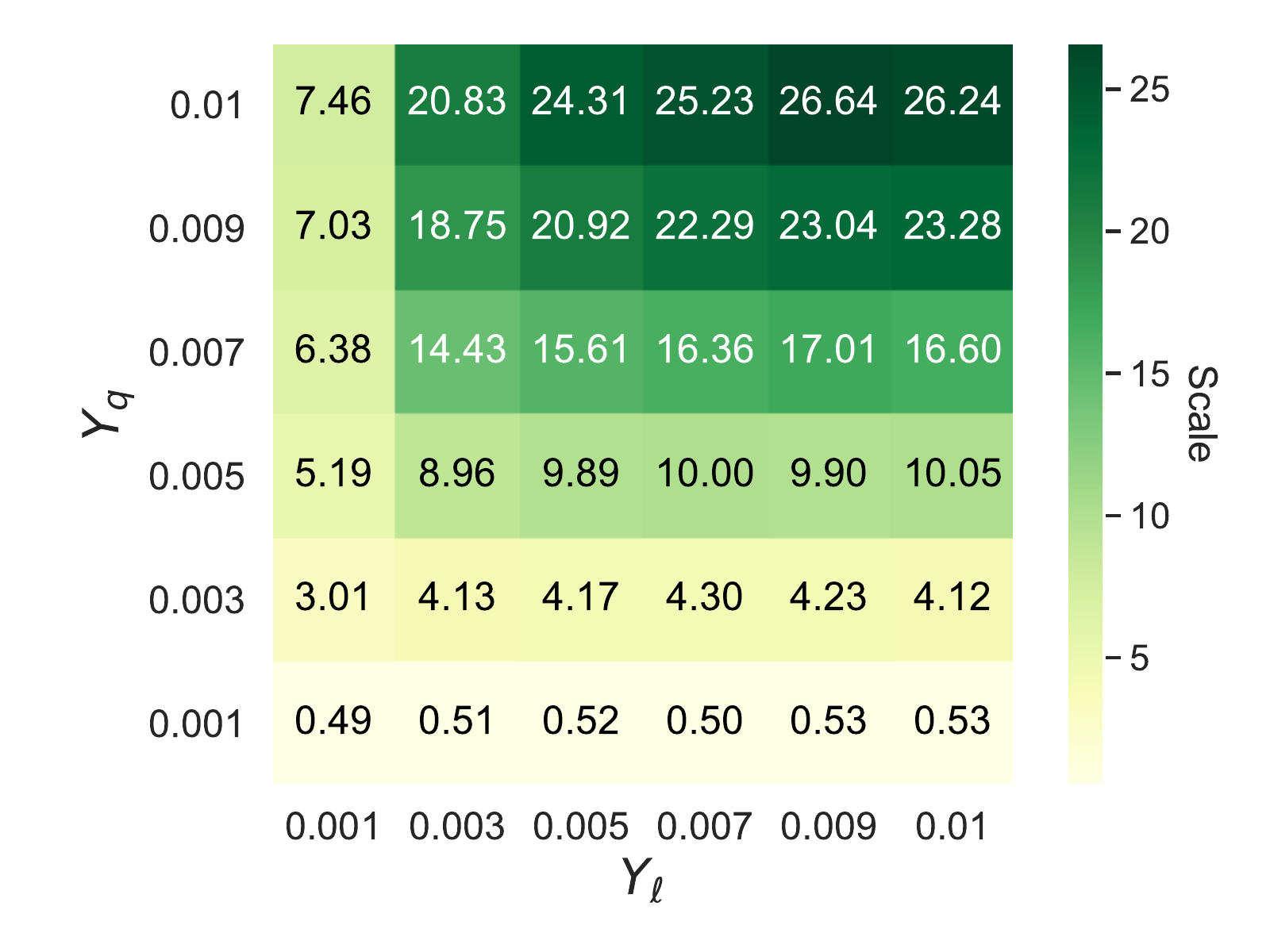}
}
\subfloat[]{
  \label{fig:cut-sign-20-170}
	\centering
  \includegraphics[width=0.49\textwidth]{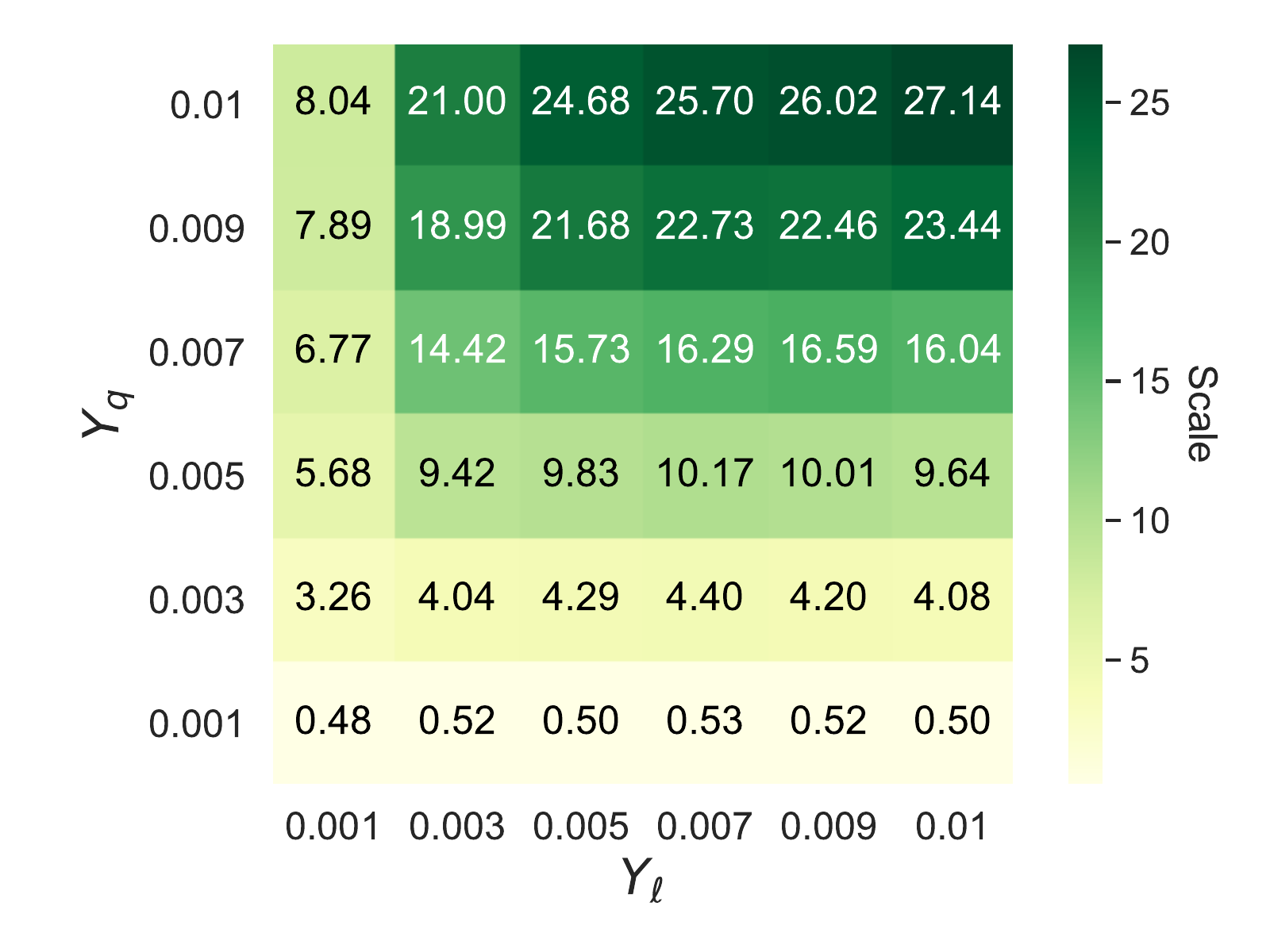}
}
\hspace{0.2\textwidth}
\centering
\subfloat[]{
  \label{fig:cut-sign-60-160}
	\centering
  \includegraphics[width=0.49\textwidth]{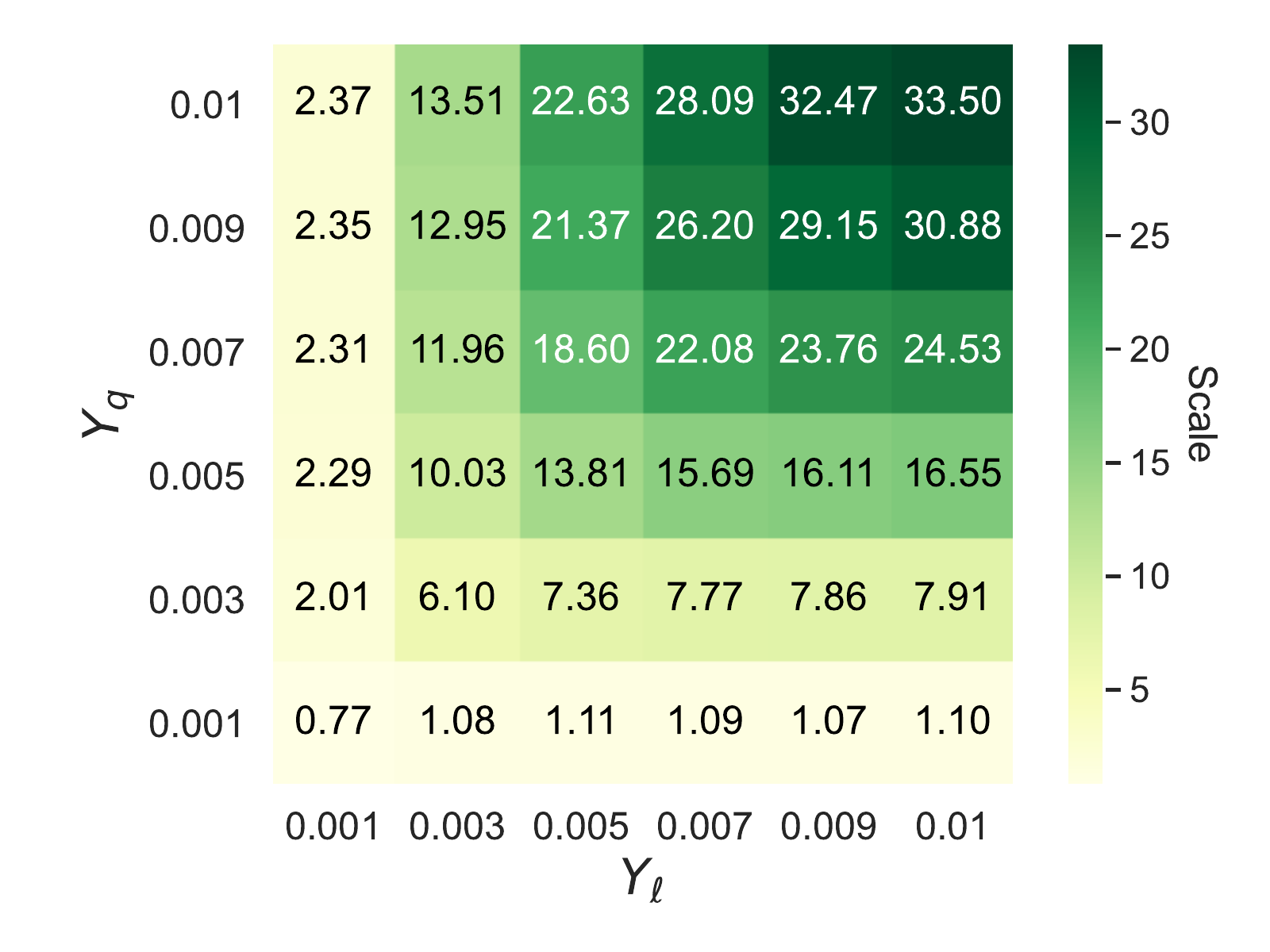}
}
\subfloat[]{
  \label{fig:cut-sign-60-170}
	\centering
  \includegraphics[width=0.49\textwidth]{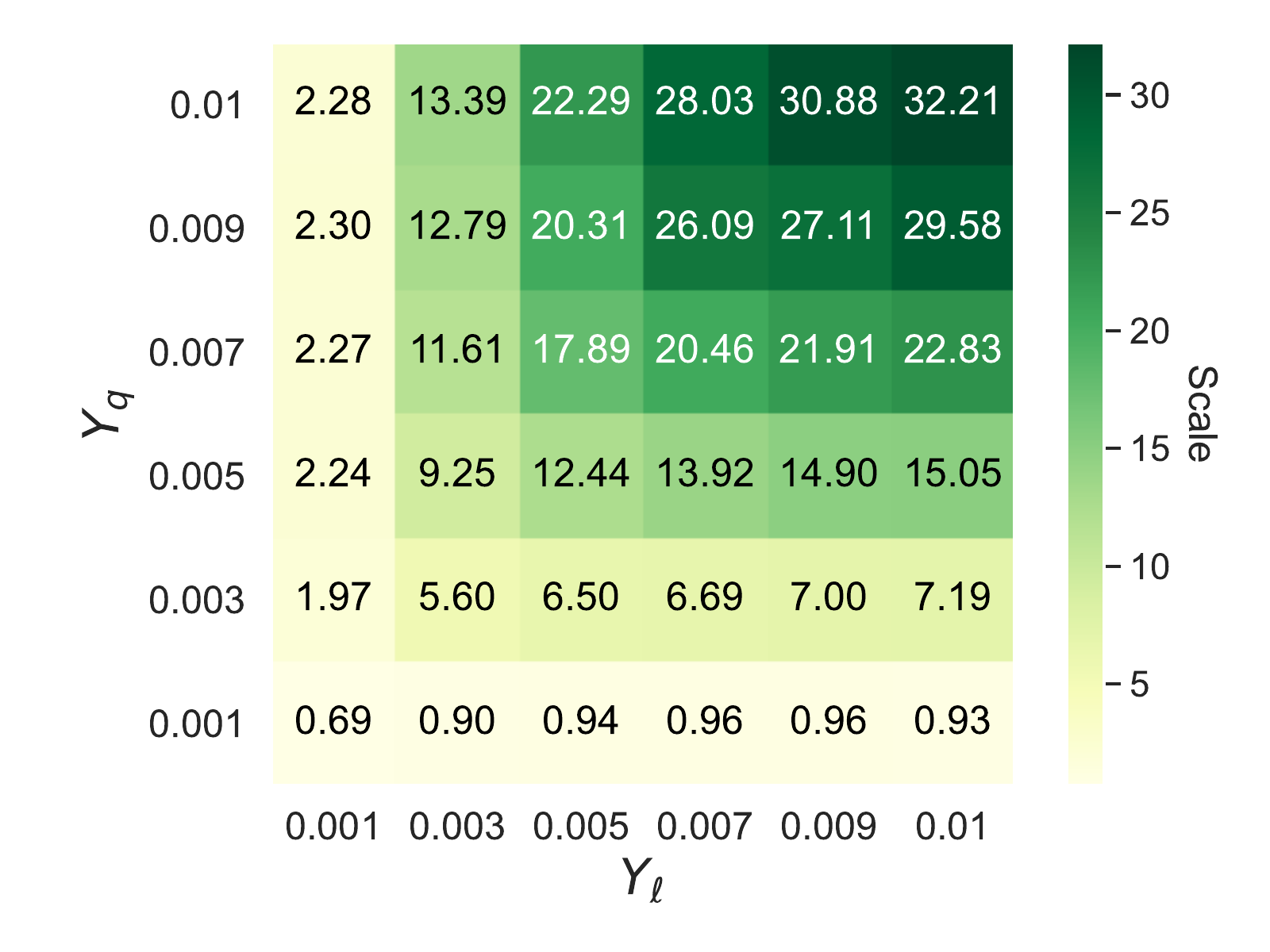}
}
\caption{\small \em Significance plots for the four mass benchmark
  configuration points at ${\cal L} = 300 \,{fb}^{-1}$ following the
  cut based analysis. The two plots in the upper panel are for
  $(m_{\chi}, m_{H})$ in GeV : (a) $(20,\,160)$, (b) $(20,\,170)$. The
  lower panel contains the other two mass points, in GeV, (c) $(60,\,160)$ and
  (d) $(60,\,170)$.}
\label{fig:cutBasedSign}
\end{figure}

In Table~\ref{tab:cutYield} we display the effects of $\met$ cut on
baseline selected signal and background events, where the last column
shows the number of signal and background events at ${\cal L} =
300~{\rm fb^{-1}}$. For signal events, we select a few representative
benchmark points to show the efficiencies of cut based analysis. In the
final signal significance calculation, we use all the signal
benchmarks points that are shown in Section~\ref{sec:bps}.

In Figure~\ref{fig:cutBasedSign}, the panels $(a)$, $(b)$, $(c)$ and
$(d)$ represent signal significances in the $Y_\ell -Y_q$ Yukawa plane
at 14 TeV LHC with ${\cal L}=300~{\rm fb}^{-1}$ corresponding to
$(m_\chi, m_H)$ = (20, 160), (20, 170), (60, 160) and (60, 170) GeV,
respectively.  As can be inferred from the plots, for a given value of
$m_\chi$ and $m_H$, the significance increases with increase in
$Y_\ell$ and $Y_q$. The reason is easy to understand from the
functional dependence of the signal cross section on $Y_\ell$ and
$Y_q$ as showcased in Figure~\ref{fig:xsec}.
The cut based analysis shows that for all of these four mass
benchmark points, the signal significance ${\cal S} > 5\sigma$ is achievable
for $Y_\ell~(Y_q)$ as low as $0.001~(0.003)$. By proper scaling one can easily
obtain the signal significance at higher luminosities. For example, by looking
at Figure~\ref{fig:cutBasedSign} for $(Y_\ell, Y_q)=(0.001, 0.001)$, we find
though that the required luminosity for $5\sigma$ significance is above
$3000~{\rm fb^{-1}}$ which is beyond the reach of HL-LHC. For the benchmarks with
lighter pseudoscalar as shown in Figures \ref{fig:cut-sign-20-160} and
\ref{fig:cut-sign-20-170}, we find that the required luminosity for $5\sigma$
significance for $(Y_\ell,\,Y_q)\,=\,(0.001,\,0.003)$ is ${\cal L}_{5\sigma} \sim 800~{\rm fb}^{-1}$.
Similarly, for heavier mass of $\chi$ shown in Figures \ref{fig:cut-sign-60-160}
and \ref{fig:cut-sign-60-170}, we require ${\cal L}_{5\sigma} \sim 1900~{\rm fb}^{-1}$
integrated luminosity to achieve a $5\sigma$ significance with the same set of couplings.
These clearly indicate that the high luminosity
option of the LHC has enough potential to dig out such exotically behaving
spin-$0$ states.  Here we point out that for a few values of $Y_\ell$
and $Y_q$ in Figure~\ref{fig:cutBasedSign}, the value of the
significance ${\cal S}$ remains either the same or gets smaller for
the next higher value of either $Y_\ell$ or $Y_q$. These anomalies,
however, are the results of statistical fluctuation while estimating
the signal significance.  Also, the signal significance has been
estimated using simple cut based analysis where we have considered
real physics backgrounds only, neglecting various fake rates and
systematic uncertainties associated with various SM background
estimations. Hence, these significance values may be considered merely
as indicative ones.


\subsection{Signal extraction : Multivariate analysis} \label{mvaanalysis}

It emerged in recent years that the application of multivariate
analysis\,(MVA) provided better separation between the signal and
background than the usual rectangular cut based
analysis~\cite{Bardhan:2016gui,Bhardwaj:2019mts,Konar:2021nkk,CMS:2011oen,Albertsson:2018maf,CMS:2018sxu}.
Inspired by these studies, we proceed to perform the multivariate
analysis using the BDT~\cite{QUINLAN1987221}
algorithm to explore the possibility of improving the signal
significance over the cut based one. In Table~\ref{tab:BDT_variables}
we show the list of input variables used for the training and
validation of our BDTs. After the baseline selection, we have trained
BDTs separately for the four signal mass points mentioned earlier. We
use a simple BDT architecture in the Root TMVA~\cite{Hocker:2007ht}
package with the set of parameters as shown in
Table~\ref{tab:BDT_parameters}.

\begin{table}[!h]
\begin{center}
\begin{tabular}{|l|c|}
\hline
Variable & Definition\\
\hline
\texttt{$p_{T}^{\mu^\prime}$} & Transverse momentum of $\mu^\prime$  \\
\texttt{$p_{T}^{\tau_{h}}$} & Transverse momentum of the $\tau_{h}$ from $\chi$ \\
\texttt{$\met$} & Missing transverse energy \\
\texttt{$p_T^Z$} & Transverse momentum of the selected $Z$ candidate \\
\texttt{$\Delta R_{l^{+}l^{-}}$} &  $\Delta R$ between the OSSF leptons from $Z$ decay \\
\texttt{$m_{T}(\met,\, e/\mu)$} & Transverse mass of $\met$ and $\mu^{\prime}$ \\
\multirow{2}{*}\texttt{$\alpha_{\chi, Z}$} & Angle between the planes
of the pair of $\mu^{\prime}$, $\tau_{h}$ and the pair consisting \\ &
of two OSSF leptons from $Z$ \\

\hline
\end{tabular}
\end{center}
\caption{\small \em List of kinematic variables used in the BDT based analysis.}
\label{tab:BDT_variables}
\end{table}

\begin{table}[!h]
  \begin{center}
    \footnotesize
\begin{tabular}{|l|c|c|}
\hline
BDT parameters & Description & Value\\
\hline
\texttt{NTrees} & Number of trees or nodes & $750$ \\
\texttt{MinNodeSize} & Minimum $\%$ of training events required in a leaf node & $5 \%$\\
\texttt{MaxDepth} & Max depth of the decision tree allowed & $3$\\
\texttt{BoostType} & Boosting mechanism to make the classifier robust & AdaBoost~\cite{FREUND1997119} \\
\texttt{AdaBoostBeta} & Learning rate for AdaBoost algorithm & $0.5$\\
\multirow{2}{*}{\texttt{nCuts}} & Number of grid points in variable range & \\
                                & used in finding optimal cut in node splitting & $20$ \\
\hline
\end{tabular}
\end{center}
\caption{\small \em The list of BDT parameters : definition and values used.}
\label{tab:BDT_parameters}
\end{table}

\begin{figure}[!h]
\centering
\subfloat[]{
  \label{fig:BDT_20_160_score}
	\centering
  \includegraphics[width=0.45\textwidth]{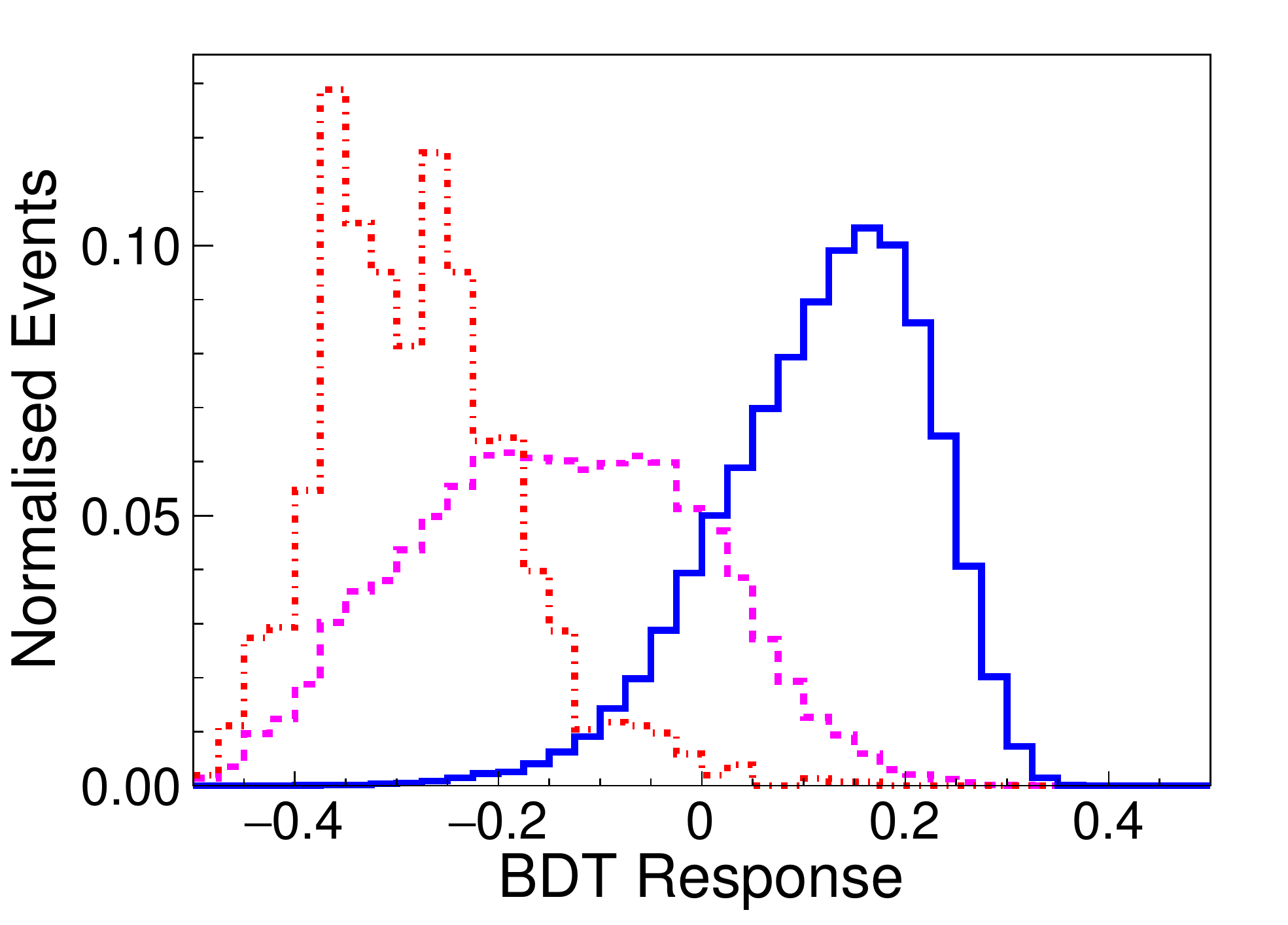}
}
\subfloat[]{
  \label{fig:BDT_20_170_score}
  \centering
  \includegraphics[width=0.45\textwidth]{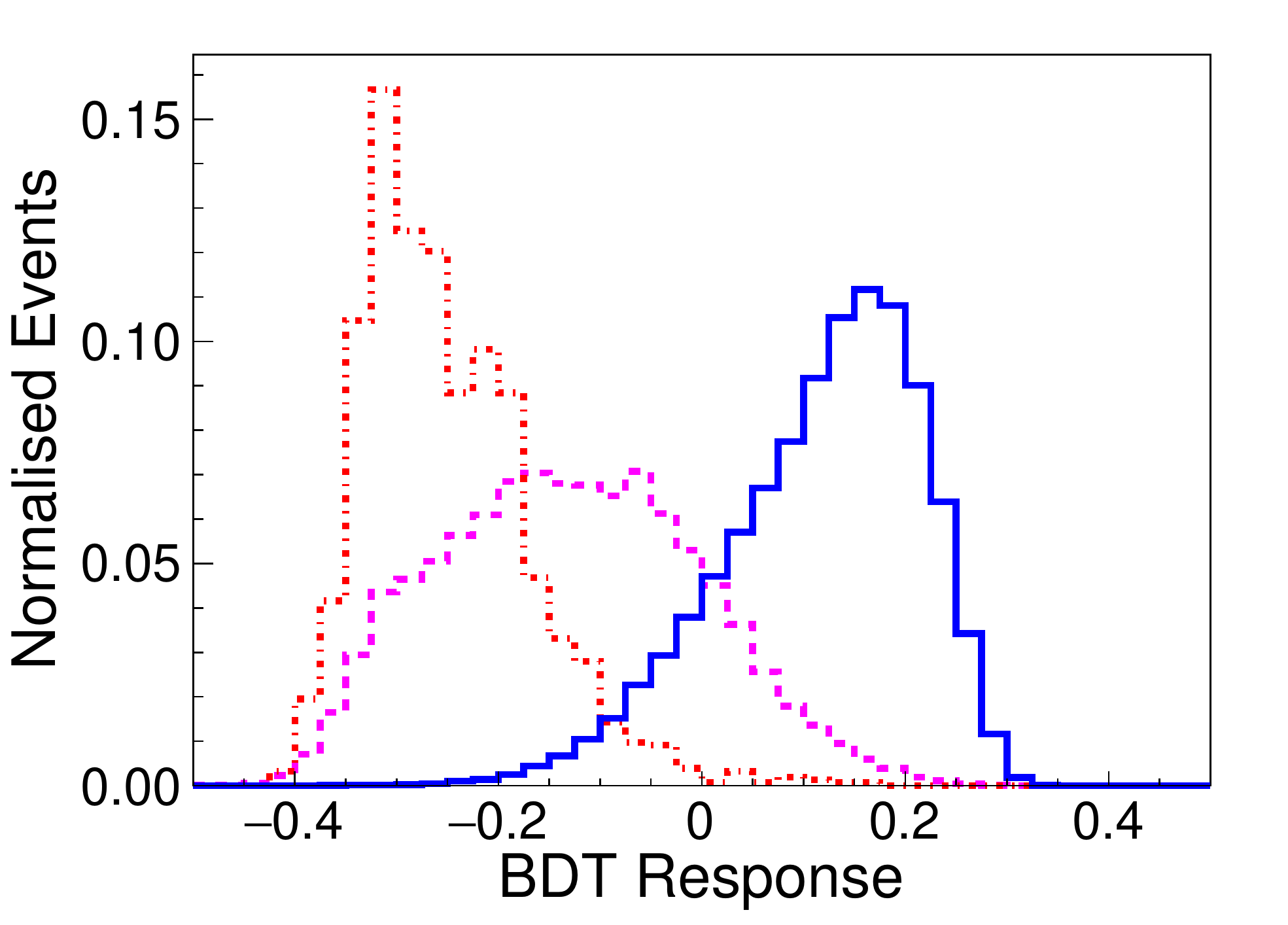}
}
\hspace{0.2\textwidth}
\centering
\subfloat[]{
  \label{fig:BDT_60_160_score}
  \centering
  \includegraphics[width=0.45\textwidth]{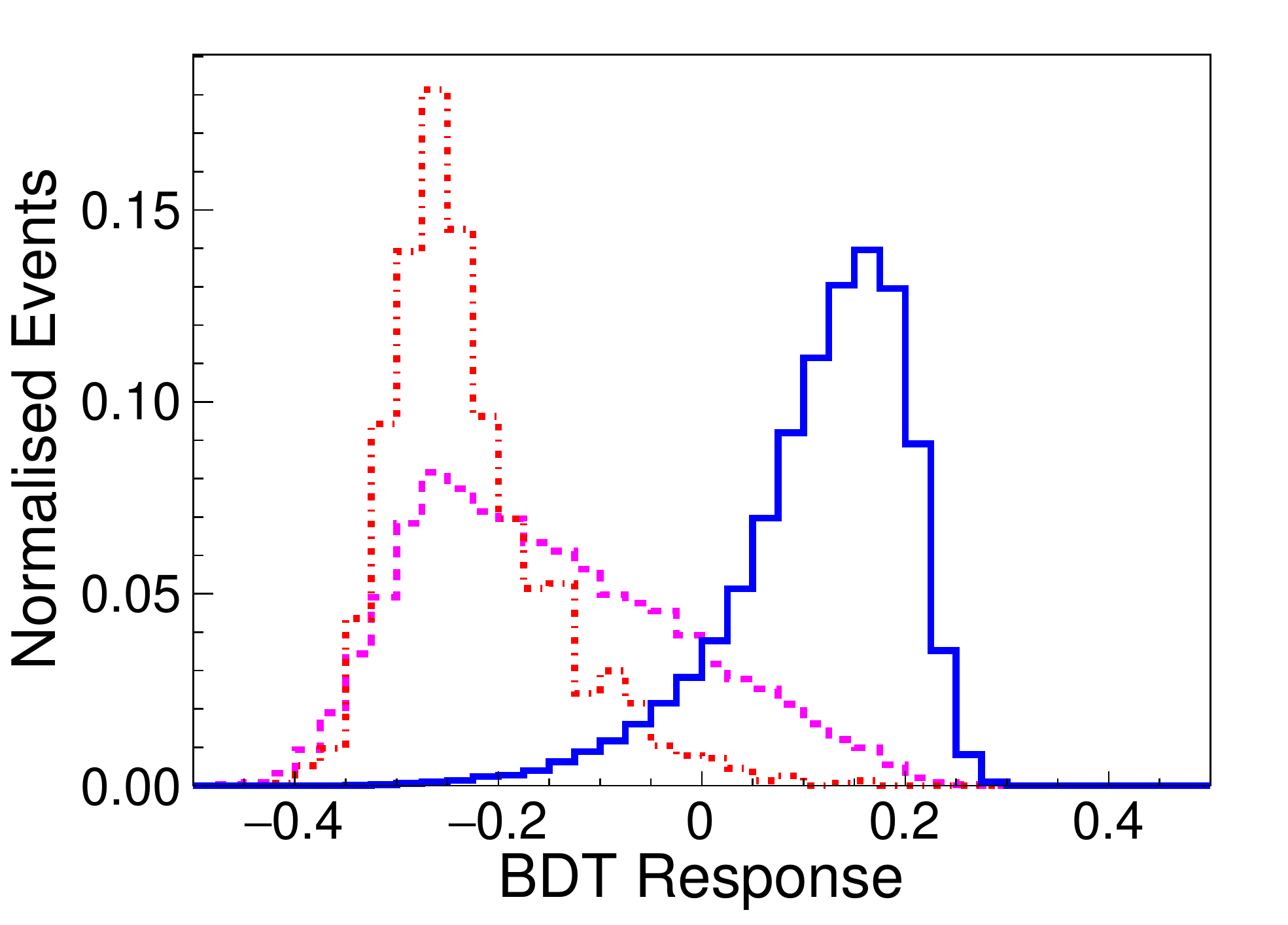}
}
\subfloat[]{
  \label{fig:BDT_60_170_score}
  \centering
  \includegraphics[width=0.45\textwidth]{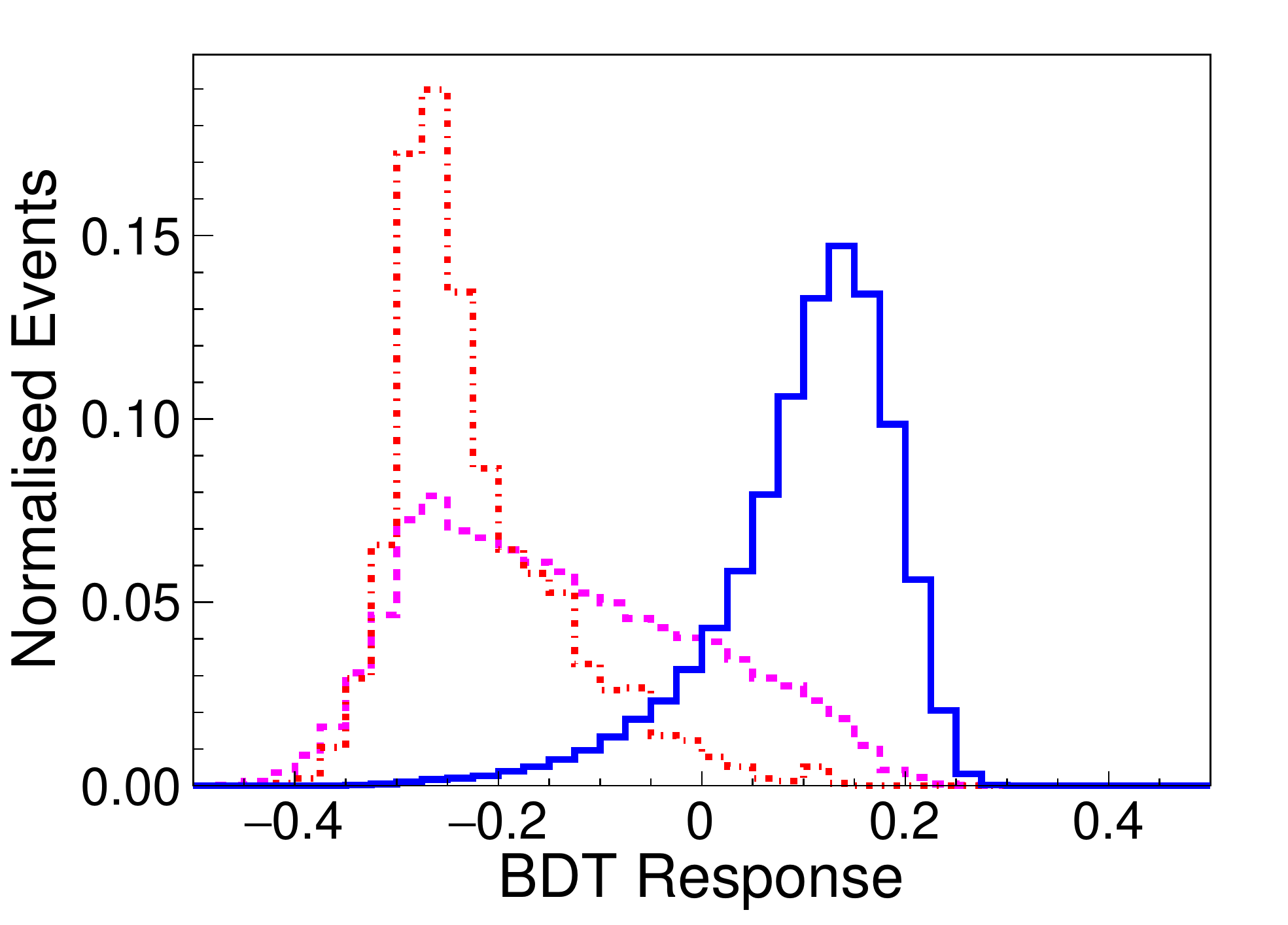}
}
  \label{fig:legend2}
  \centering
  \includegraphics[width=0.8\textwidth]{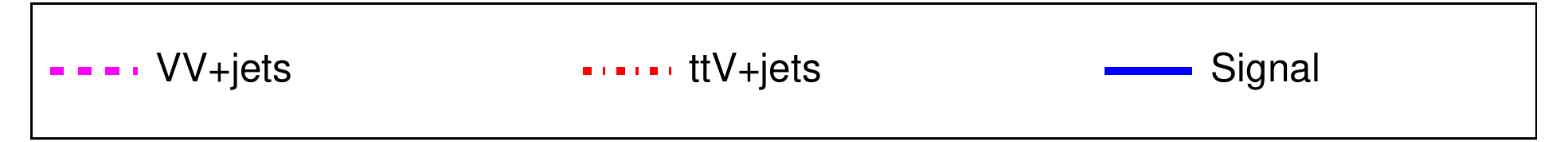}
\caption{\small \em Shape of the BDT discriminants for signal and
  backgrounds. The four plots are for four mass benchmarks:
  $(m_{\chi}, m_{H})$ in GeV $=$ (a) $(20,\,160)$, (b) $(20,\,170)$, (c)
  $(60,\,160)$ and (d) $(60,\,170)$.}
\label{fig:BDTscores}
\end{figure}

\begin{table}[h!]
  \begin{center}
    {\footnotesize
      \centering
      \setlength{\tabcolsep}{0.5em} 
                {\renewcommand{\arraystretch}{1.3}
      \begin{tabular}{|>{}c|*{4}{c|}}\hline 
        \multirow{2}{*}{Samples} & \multirow{2}{*}{Baseline} & \multirow{2}{*}{Effective cross sections (fb)} & \multirow{2}{*}{Events} \\
        &      cross section (fb)   &         passing the best BDT score             & ${\cal L}=300\, {\rm fb}^{-1}$ \\
        \hline
        \texttt{$(20,160)-(0.005,0.005)$}     & $0.845$ & $0.453$ & $135.98$ \\
        \texttt{$t \bar t Z + {\rm jets}$}    & $0.141$ & $0.001$ & $0.11$   \\
        \texttt{$W^{\pm}Z + {\rm jets (3l)}$} & $1.144$ & $0.003$ & $0.91$   \\
        \texttt{$ZZ + {\rm jets ~(4l)}$}      & $1.252$ & $0.034$ & $10.21$  \\
        \hline\hline\hline
        \texttt{$(60,170)-(0.005,0.005)$}     & $1.142$ & $0.916$ & $274.97$ \\
        \texttt{$t \bar t Z + {\rm jets}$}    & $0.141$ & $0.001$ & $0.42$   \\
        \texttt{$W^{\pm}Z + {\rm jets (3l)}$} & $1.144$ & $0.031$ & $9.39$   \\
        \texttt{$ZZ + {\rm jets ~(4l)}$}      & $1.252$ & $0.173$ & $52.06$  \\
        \hline
    \end{tabular}}}
  \end{center}
  \caption{\small \em Summary of multivariate analysis (notations
    for the signals same as in Table~\ref{tab:baselineYield}). Here
    we show the effect of BDT only for one coupling benchmark point
    $(Y_\ell, Y_q) = (0.005, 0.005)$.}
  \label{tab:bdtYield}
\end{table}

\begin{figure}[!h]
\centering
\subfloat[]{
  \label{fig:BDT_20_160_ROC}
	\centering
  \includegraphics[width=0.49\textwidth]{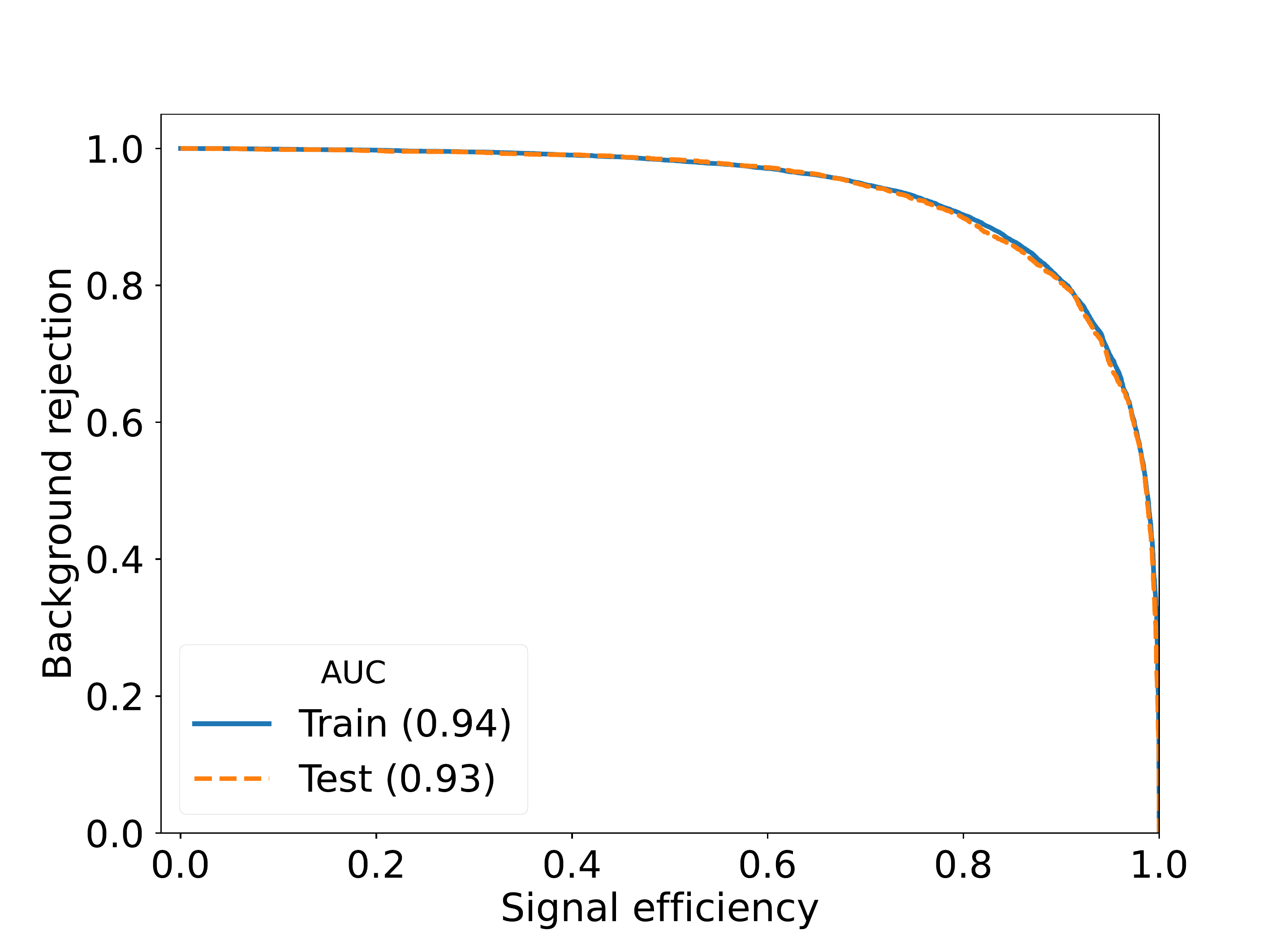}
}
\subfloat[]{
  \label{fig:BDT_60_170_ROC}
  \centering
  \includegraphics[width=0.49\textwidth]{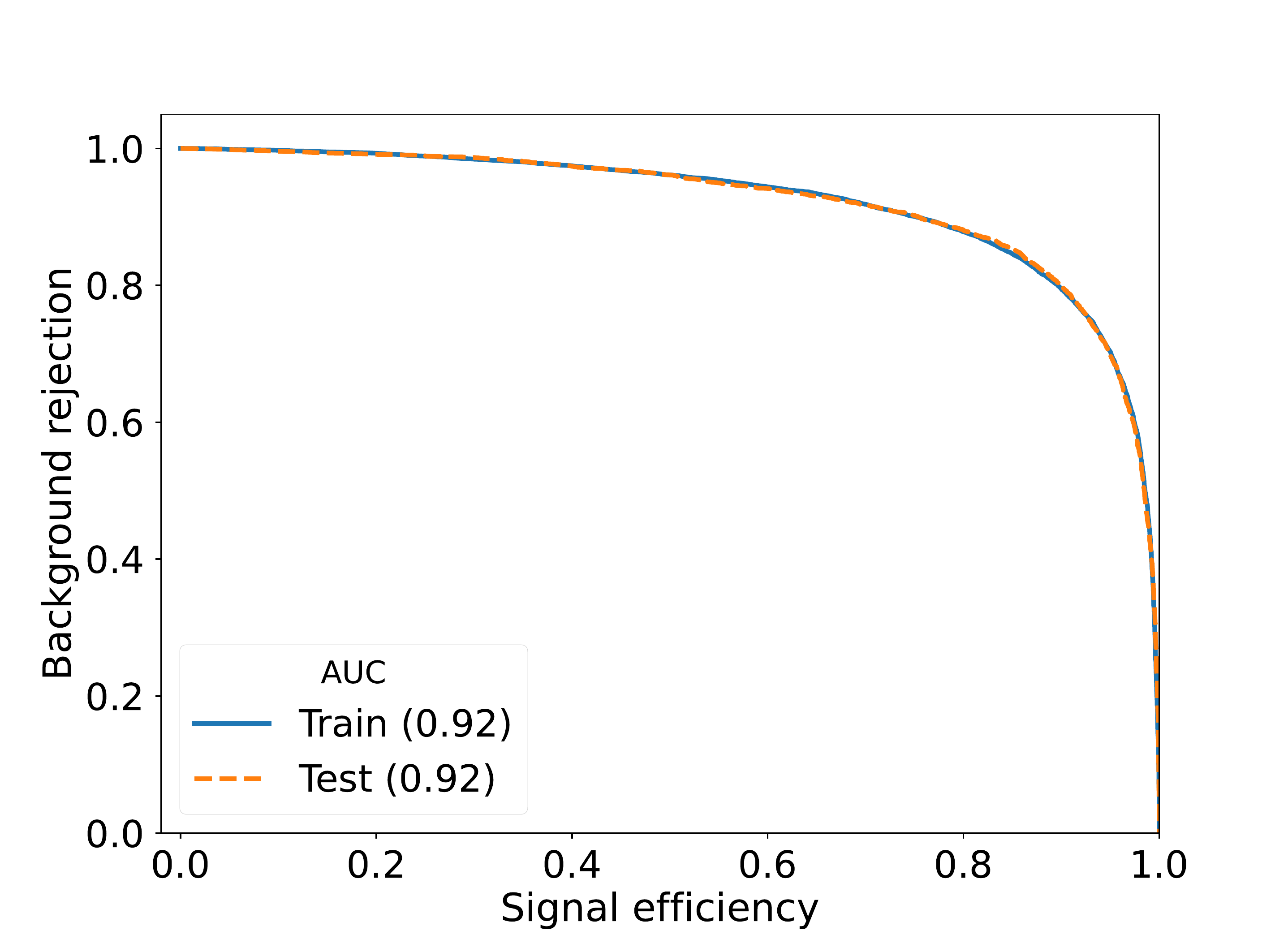}
}
\caption{\small \em ROC curves for $(m_\chi,m_H)$ in GeV: (a) $(20,160)$
  and (b) $(60,170)$ with unweighted signal and background events with their
  AUC values.}
\label{fig:ROC}
\end{figure}

For training and testing of BDTs at each mass configuration, we
combine the signal events for all the $(Y_{\ell},Y_q)$ points because
there are negligible changes in the shape of the kinematic variables
for different combinations of coupling values.  We employ a comparable
number of signal and background events (about 75\% of the latter) to
prepare the training dataset, where all the events have been selected
randomly. We only consider the two major backgrounds $W^{\pm}Z (3l) + ~{\rm jets}$
and $ZZ (4l) + ~{\rm jets}$ as shown in Table~\ref{tab:baselineYield}.
We tune the BDT parameters in Table~\ref{tab:BDT_parameters} to minimize
the over training for optimal performance.

In Figure~\ref{fig:BDTscores}~we display the distributions of the BDT
classifier score for the signal benchmark points $(m_\chi, m_H)$: (a)
(20, 160), (b) (20, 170), (c) (60,160) and (d) (60,170) GeV,
respectively, and two SM backgrounds.
We can see from Figure~\ref{fig:BDTscores} that for signal benchmarks with both lighter
(upper panels) and heavier $\chi$ (lower panels), the
classifier shows similar performance. The performance of training and the
possibility of over fitting can be inferred from the shape of the
receiver operating characteristic\,(ROC) curves as displayed in
Figure~\ref{fig:ROC}. From this figure, it is evident that the ROC
looks similar for both the training and test samples, which implies
very negligible over training in our BDT analysis. The area under the curve
(AUC) is a metric to show the performance of a classifier. A complete separation
between signal and background would make AUC$\,=\,1$ and here the value is almost
$0.93$. So, the performance of BDT and negligible overfitting undoubtedly give
us the confidence that we proceed further to estimate the final
signal significance with ${\cal L} = 300~{\rm fb}^{-1}$.

\begin{figure}[!h]
\centering
\subfloat[]{
  \label{fig:bdt-sign-20-160}
  \centering
  \includegraphics[width=0.49\textwidth]{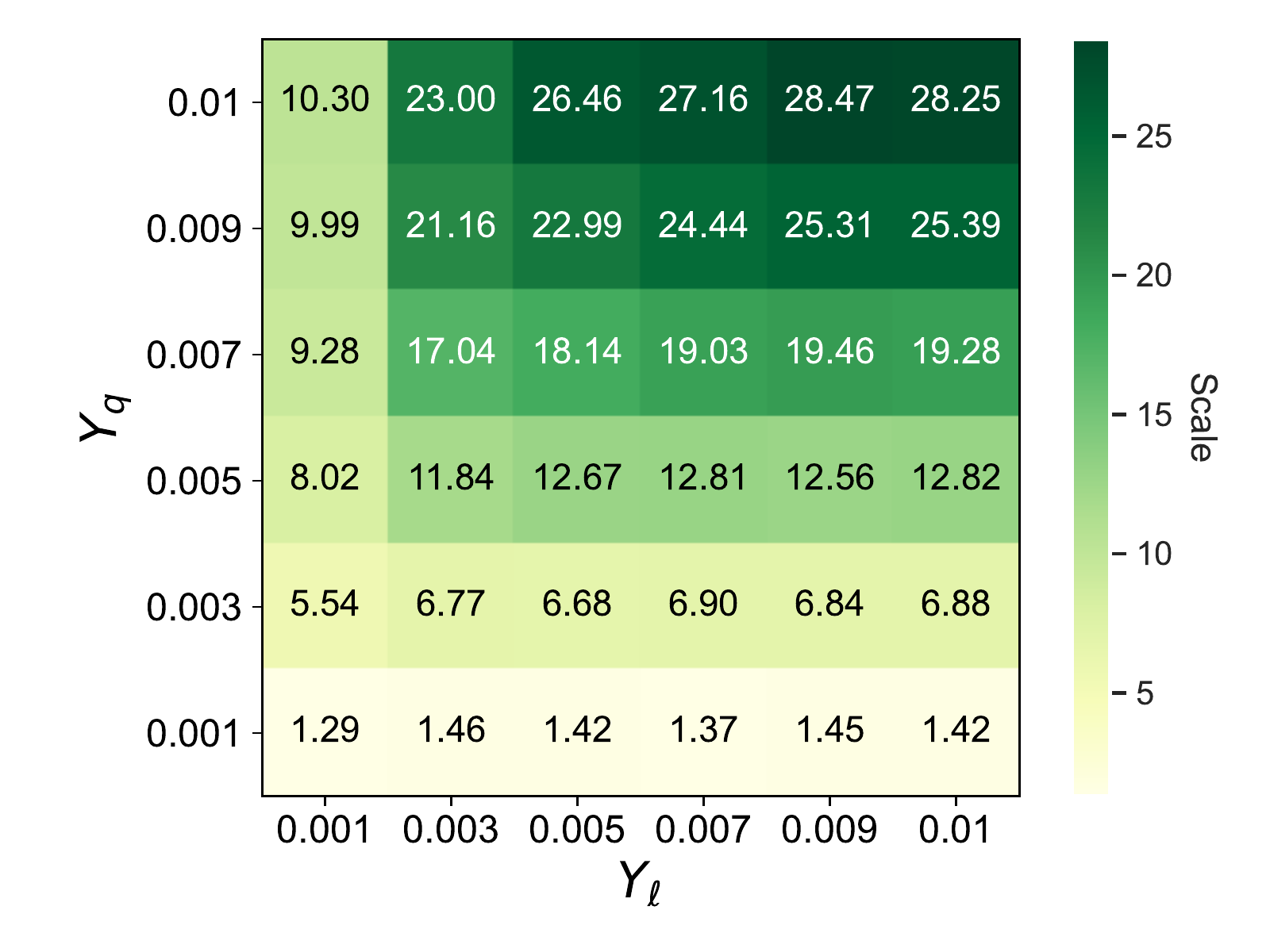}
}
\subfloat[]{
  \label{fig:bdt-sign-60-160}
  \centering
  \includegraphics[width=0.49\textwidth]{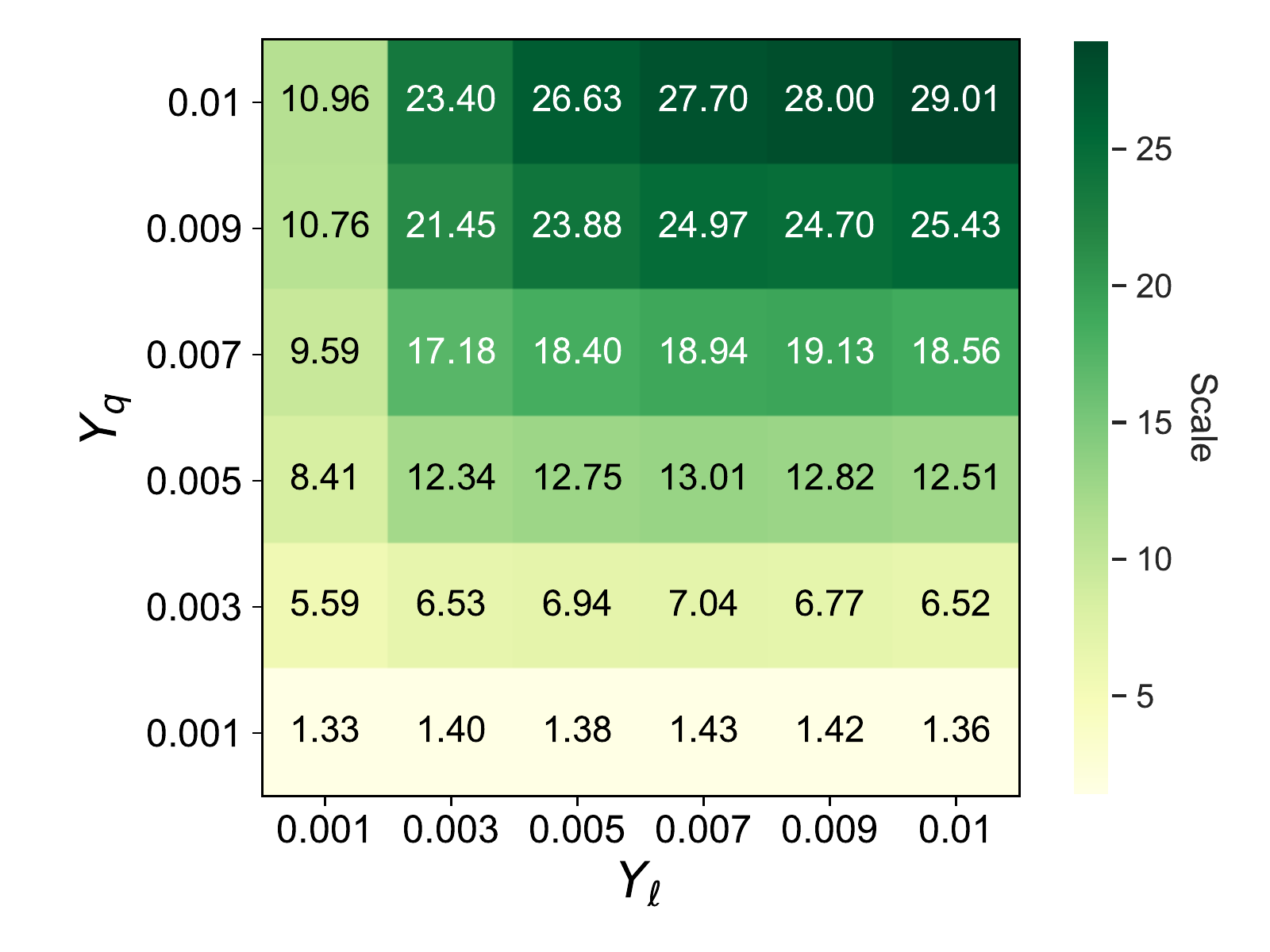}
}
\hspace{0.2\textwidth}
\centering
\subfloat[]{
  \label{fig:bdt-sign-20-170}
  \centering
  \includegraphics[width=0.49\textwidth]{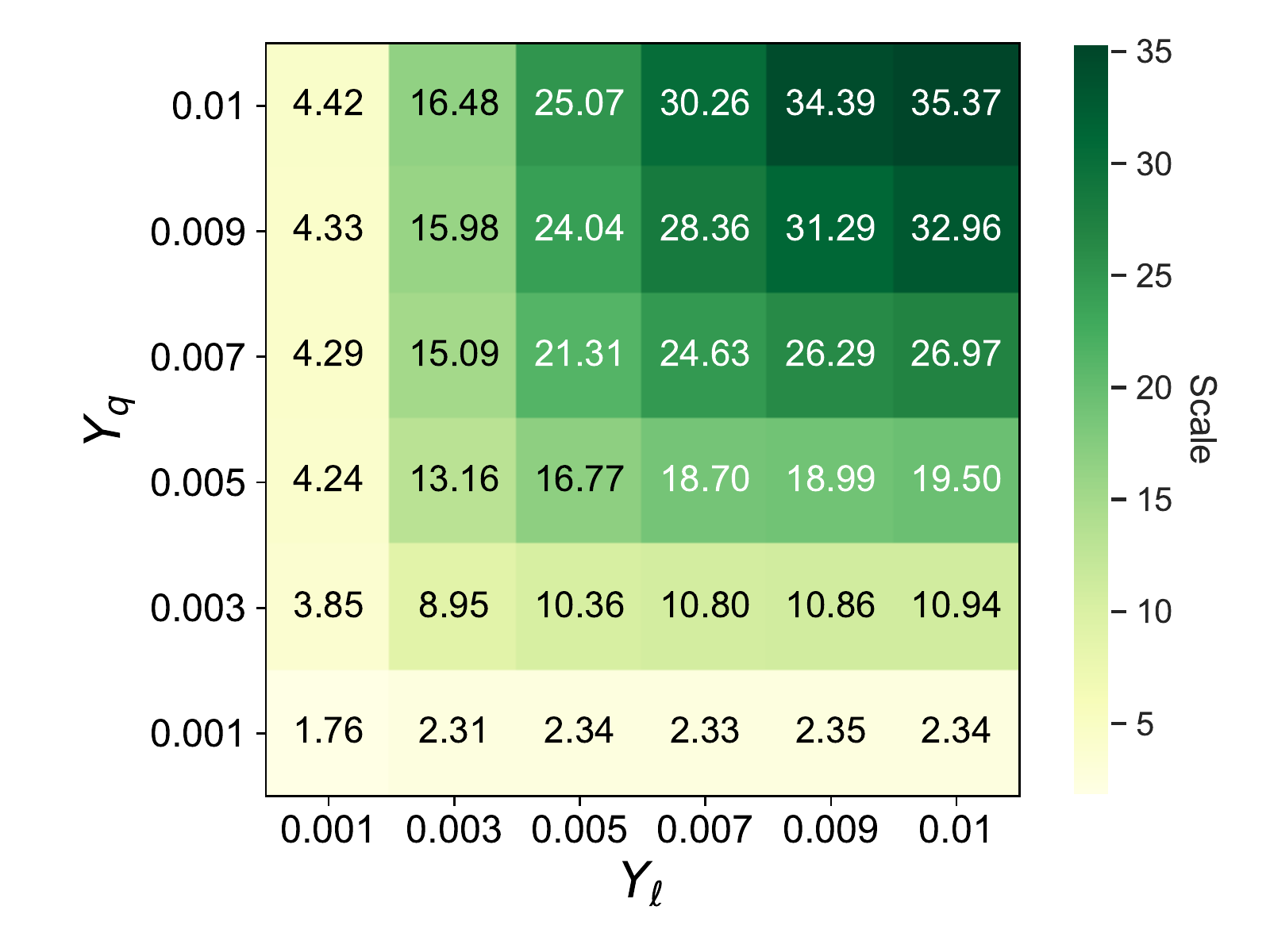}
}
\subfloat[]{
  \label{fig:bdt-sign-60-170}
  \centering
  \includegraphics[width=0.49\textwidth]{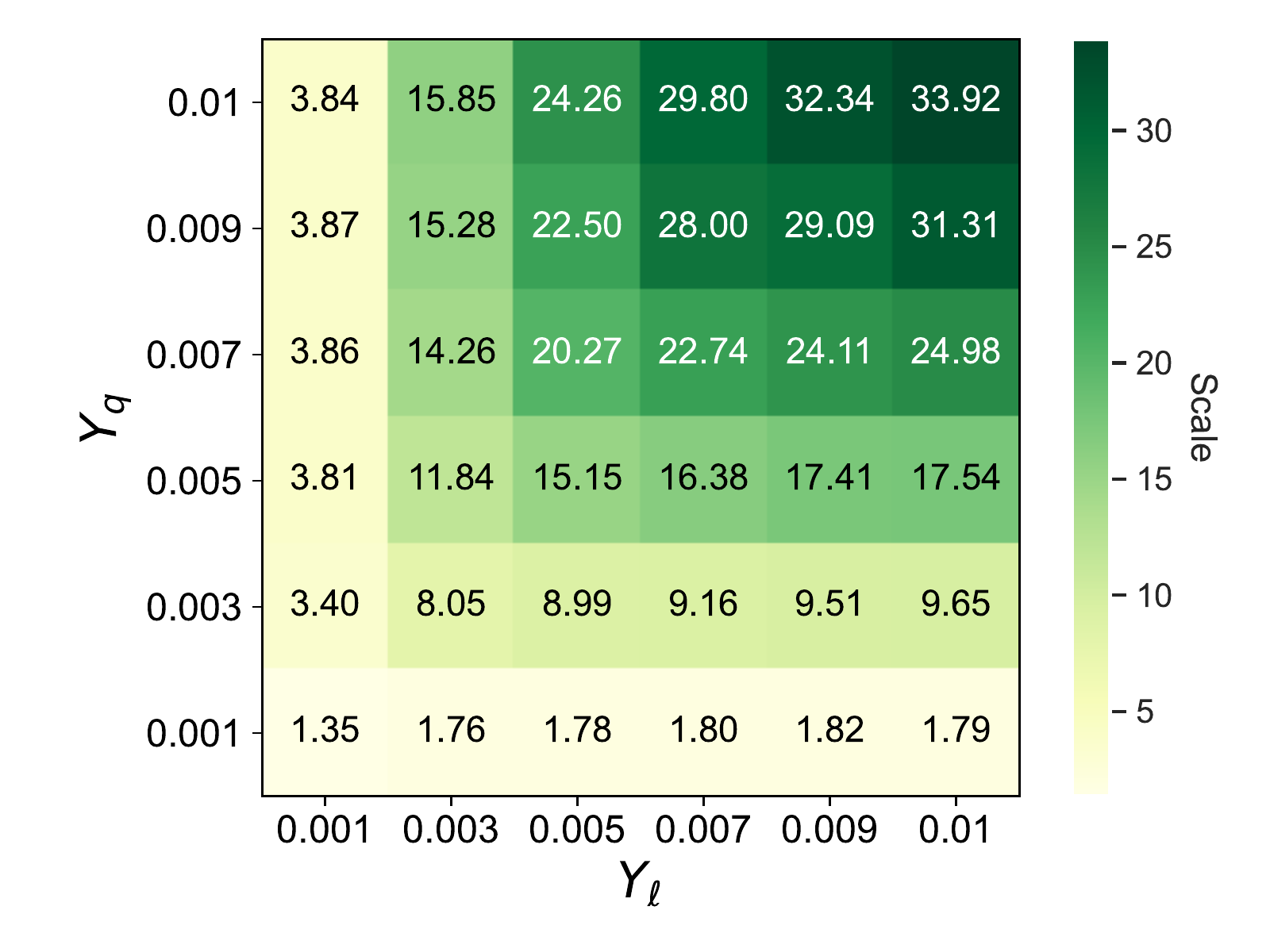}
}
\caption{\small \em Significance plots for the benchmark mass
  configurations at ${\cal L} = 300 \,{fb}^{-1}$ after BDT. The four
  plots are for $(m_\chi,m_H)$ in GeV: (a) $(20,160)$, (b) $(20,170)$, (c)
  $(60,160)$ and (d) $(60,170)$.}
\label{fig:bdtBasedSign}
\end{figure}

To get the final significance for all the signal benchmarks we first
scale the number of signal and background events with a factor of
$\sigma_{S(B)}$ $\times$ ${\cal L}$ $\times$ $\epsilon_{S(B)}$ and
then we iterate over the BDT scores to get the optimal point
corresponding to the maximum significance. In Table~\ref{tab:bdtYield}
we only show the performance of two BDT models trained for ($m_\chi,
m_H$) : (20,160) and (60, 170) GeV signal processes, respectively.  We
only tabulate the application of the BDT models on $(Y_\ell, Y_q) =
(0.005, 0.005)$ signal coupling benchmark point. The last column of
Table~\ref{tab:bdtYield} shows the final yield at ${\cal L} = 300~{\rm fb}^{-1}$
and the yield corresponds to the highest value of
significance for the scenario under consideration.  Keeping the same
format of Figure~\ref{fig:cutBasedSign} (cut based), we show the
variation of the signal significance $({\cal S})$ as a function of the
scalar masses $(m_\chi, m_H)$ and Yukawa couplings $(Y_\ell, Y_q)$ in
Figure~\ref{fig:bdtBasedSign}.  The functional dependence of ${\cal S}$
on model parameters $(m_\chi, m_H, Y_\ell, Y_q)$ remains the
same as in the cut based analysis. From these four panels, one can see
that even for a modest value of $Y_q =Y_\ell \sim 0.003~(0.005)$, we
get ${\cal S} \sim ~8(14)$ with $300~{\rm fb}^{-1}$ data.

\subsection{14 vs. 13.6 vs. 13 TeV} \label{Comparsion}

It is likely that during $2022$-$2024$, LHC would run at 13.6 TeV and during
that period it might collect 150 ${\rm fb}^{-1}$ data
\cite{1977855}.

\begin{table}[h!]
\begin{center}
\begin{tabular}{|l|c|c||c|c|}
\hline $\sigma$ [fb] &
\multicolumn{4}{c|}{$\sqrt{s}\, =\,  14\,(13.6){\rm\,TeV}$}
\\\hline\hline
$(m_\chi, m_H)$ & \multicolumn{2}{c||}{$(20,160)~{\rm\,GeV}$} & \multicolumn{2}{c|}{$(60,170)~{\rm\,GeV}$} \\
\hline \hline
\diagbox{$Y_\ell$}{$Y_q$} & $0.005$ & $0.009$ & $0.005$ & $0.009$\\
\hline
\texttt{$0.003$} & $61.87(59.56)$ & $158.03(152.42)$ & $28.87(27.74)$ & $42.97(41.47)$ \\
\texttt{$0.007$} & $68.35(65.83)$ & $208.77(201.03)$ & $49.34(47.64)$ & $115.89(111.31)$ \\
\hline
\end{tabular}
\end{center}
\caption{\small \em {\bf S1} cross sections $\times$ branching ratios
  for a few benchmark points at $\sqrt s\,=\, 14(13.6)$ TeV.}
\label{tab:compxsec}
\end{table}

Eventually in 2027 the center of mass energy might be
jacked up to 14 TeV and, subsequently, availability of an integrated
luminosity as high as 3000 ${\rm fb}^{-1}$ will be on the cards. In
Table~\ref{tab:compxsec} we display a comparison of {\bf S1} cross
sections $\times$ branching ratios at $\sqrt s =$ 14 TeV and 13.6 TeV
for a few benchmark points. The reduction in cross section for 13.6
TeV, as shown in Table~\ref{tab:compxsec}, is only $(3-4)\%$.

\begin{table}[h!]
  \begin{center}
    \begin{tabular}{|l|c|c||c|c|}
      \hline $\sigma[{\rm \,fb}]({\cal S})$ & \multicolumn{4}{c|}{$\sqrt{s}\, =\,13 {\rm\,TeV}$} \\\hline\hline
      $(m_\chi, m_H)$                       & \multicolumn{2}{c||}{$(20,160)~{\rm\,GeV}$} & \multicolumn{2}{c|}{$(20,170)~{\rm\,GeV}$} \\\hline \hline
      \diagbox{$Y_\ell$}{$Y_q$}             & $0.007$         & $0.01$          & $0.007$         & $0.01$          \\ \hline
      \texttt{$0.007$}                      & $118.58(12.86)$ & $230.23(18.36)$ & $98.40(12.80)$  & $191.81(18.72)$ \\ \hline
      \texttt{$0.01$}                       & $121.33(13.03)$ & $240.12(19.09)$ & $100.34(12.54)$ & $199.22(19.61)$ \\ \hline \hline
      $(m_\chi, m_H)$                       & \multicolumn{2}{c||}{$(60,160)~{\rm\,GeV}$}         & \multicolumn{2}{c|}{$(60,170)~{\rm\,GeV}$} \\\hline \hline
      \texttt{$0.007$}                      & $92.78(16.65)$  & $138.99(16.78)$ & $75.66(15.37)$  & $118.43(20.14)$ \\ \hline
      \texttt{$0.01$}                       & $107.97(18.23)$ & $179.19(23.91)$ & $87.65(16.88)$  & $150.99(22.93)$ \\ \hline
    \end{tabular}
  \end{center}
  \caption{\small \em {\bf S1} cross sections $\times$ branching ratios (significance at ${\cal L} = 137.1\,fb^{-1}$) for a few benchmark points at $\sqrt s\,=\,13$ TeV.}
  \label{tab:comp13}
\end{table}

Thus running LHC at a little bit lower energy, even keeping the same
background contributions, costs the signal significance only
marginally for the same beam luminosity. Also, the shape of the
kinematic distributions does not significantly change either.

How do the predictions of our scenario confront the existing run 2 LHC
data?  In Table~\ref{tab:comp13}, we display the signal cross sections
times branching ratios at $\sqrt{s}\, =\,13 {\rm\,TeV}$ along with the
significance for ${\cal L} = 137.1\,{\rm fb^{-1}}$ run 2 data
collected by the CMS collaboration \cite{CMS:2021rsq}.  Interestingly,
the numbers for run 2 are already quite encouraging, worth a dedicated
study by the experimental groups.

\section{Effects of systematic uncertainties with extended benchmark scenario} \label{ext}

In Section \ref{EvSelection}, we estimated the signal significance with
four representative mass configurations of $(m_\chi,\,m_H)$, namely (20, 160/170) GeV and
(60, 160/170) GeV. Using the cut based method and BDT, we
obtained promising sensitivities to probe this channel.
\begin{figure}[!tbh]
  \centering
  \includegraphics[width=0.7\textwidth]{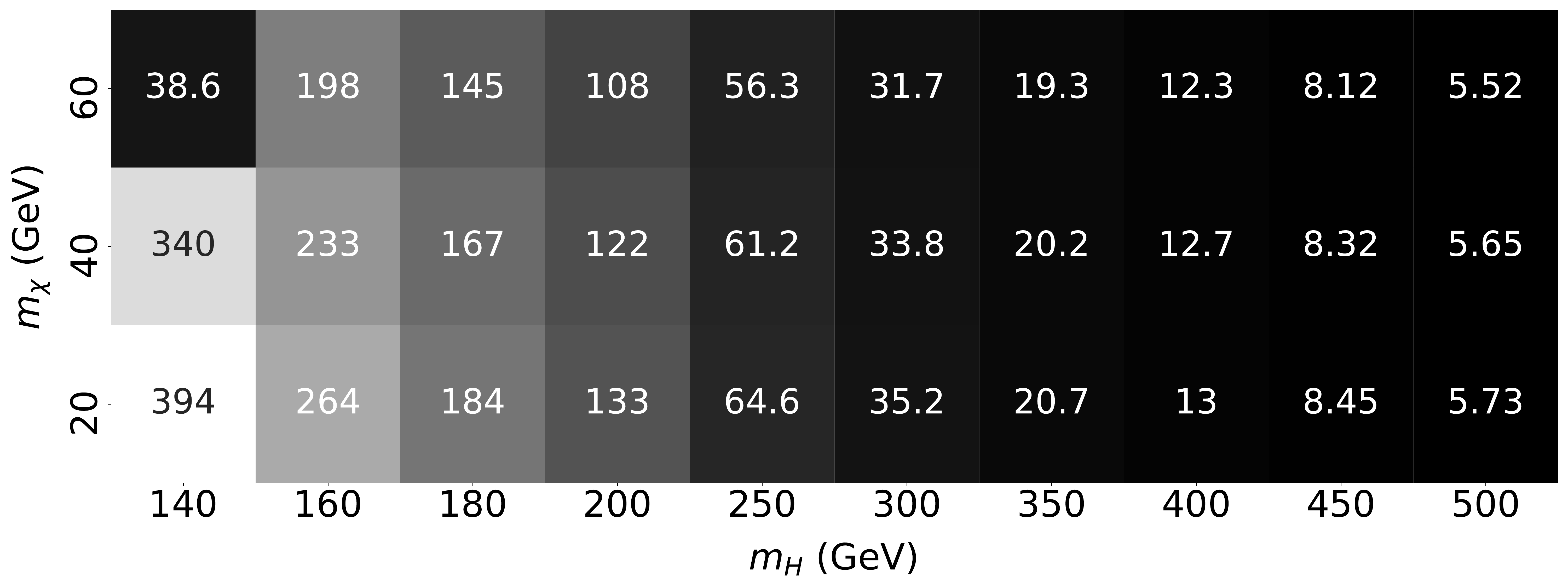}
  \caption{\small \em {\bf S1} Cross sections $\times$ branching ratios (fb) for additional mass benchmark points
    with $(Y_\ell, Y_q)\,=\,(0.01,0.01)$.}
  \label{fig:xsec_ext}
\end{figure}
In this Section, we give a broader perspective of our analysis employing an
extended benchmark region as mentioned at the end of Section \ref{sec:bps},
including possible impact of systematic uncertainties at High Luminosity (HL) option of the LHC.

In Figure \ref{fig:xsec_ext} we display the cross sections for a specific coupling combination {\it i.e.} $Y_\ell\,=\,Y_q\,=\,0.01$.
Additionally, we choose $Y_\ell\,=\,Y_q\,=\,(0.003,\,0.005,\,0.01)$ to feel the impact of possible
systematic uncertainties by varying $m_H$ in the range $140-500$ GeV and $m_\chi$ in the range $20-60$ GeV.
Clearly, the extended region thus cover the previously considered specific benchmark points analysed in Section \ref{EvSelection}.

\begin{figure}[!h]
\centering
\subfloat[]{
  \label{fig:mutaudr_LM}
  \centering
  \includegraphics[width=0.47\textwidth]{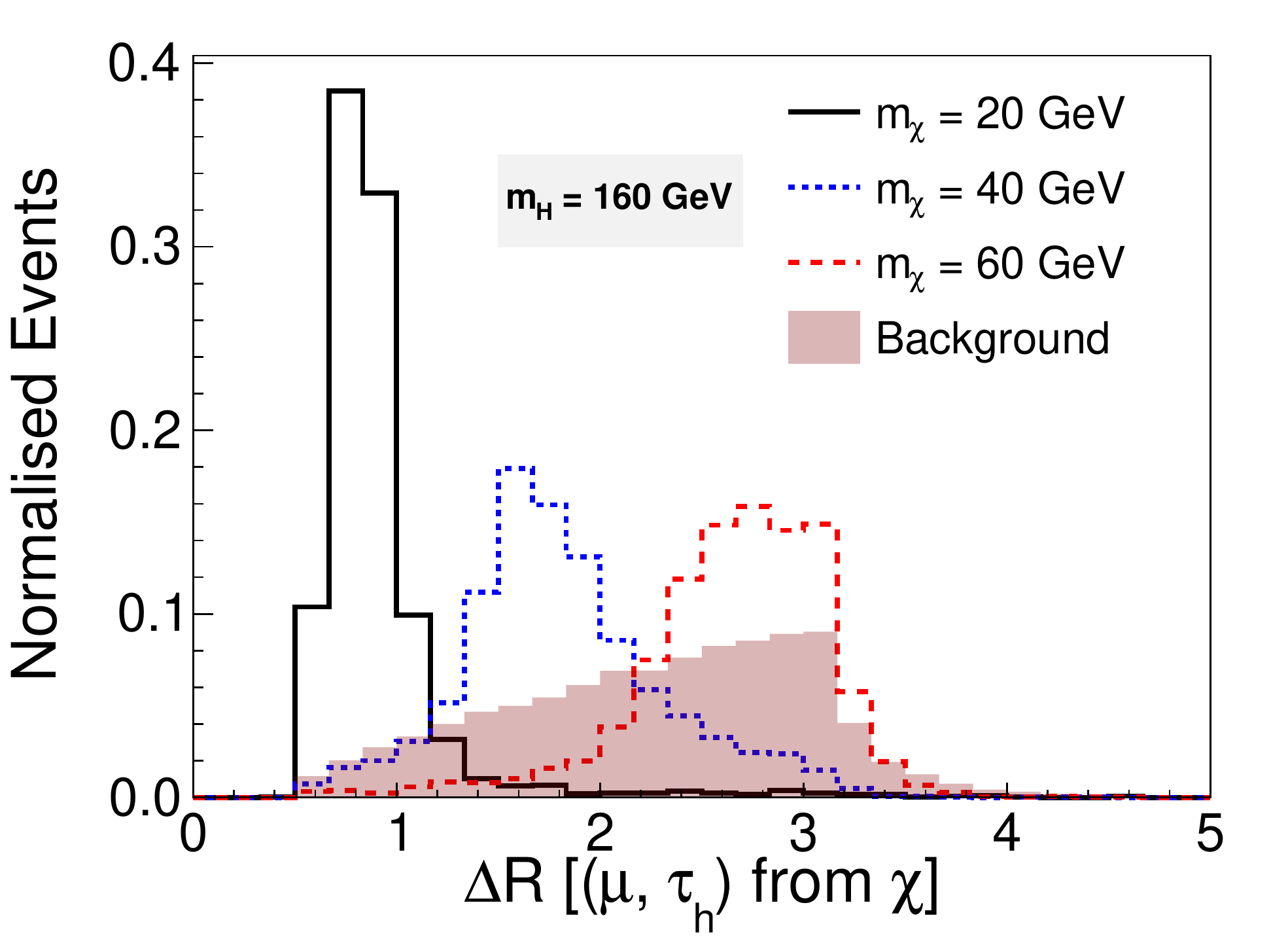}
}
\subfloat[]{
  \label{fig:mutaudr_HM}
  \centering
  \includegraphics[width=0.47\textwidth]{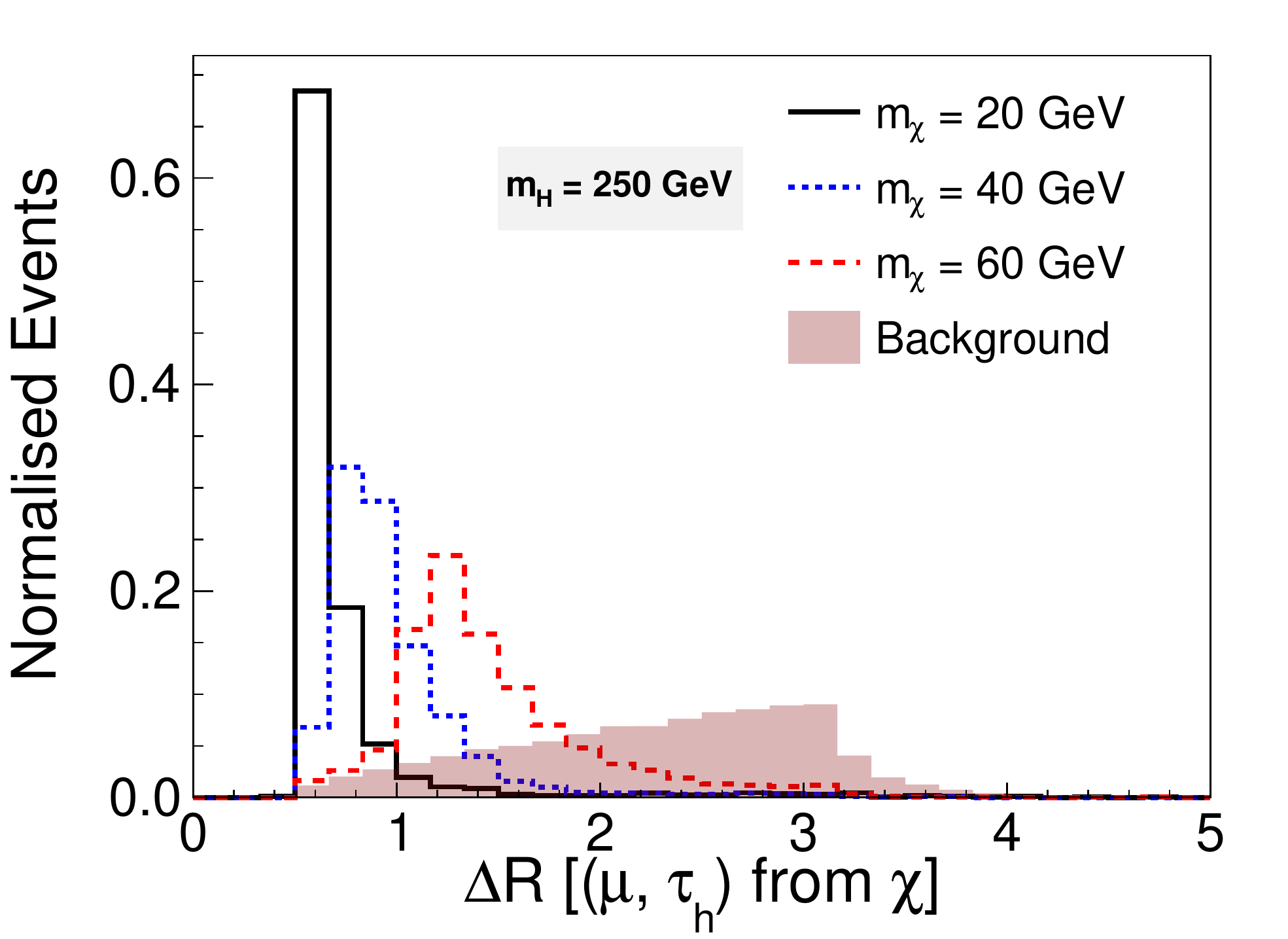}
}
\caption{\small \em $\Delta R \,(\mu^\prime,\,\tau_h)$ for
  $m_H$ (in GeV) $=$ (a) $160$, (b) $250$ with different choices of $m_\chi$.}
\label{fig:mutaudr_ext}
\end{figure}

A comparison between Figures \ref{fig:mutaudr_LM} and \ref{fig:mutaudr_HM} shows that for
higher values of $m_H$ ({\it e.g.} 250 GeV) a uniform cut on $\Delta R$ between $\mu^\prime$ and $\tau_h$
\begin{figure}[!h]
\centering
\subfloat[]{
  \label{fig:signif_0p003}
  \centering
  \includegraphics[width=0.3\textwidth]{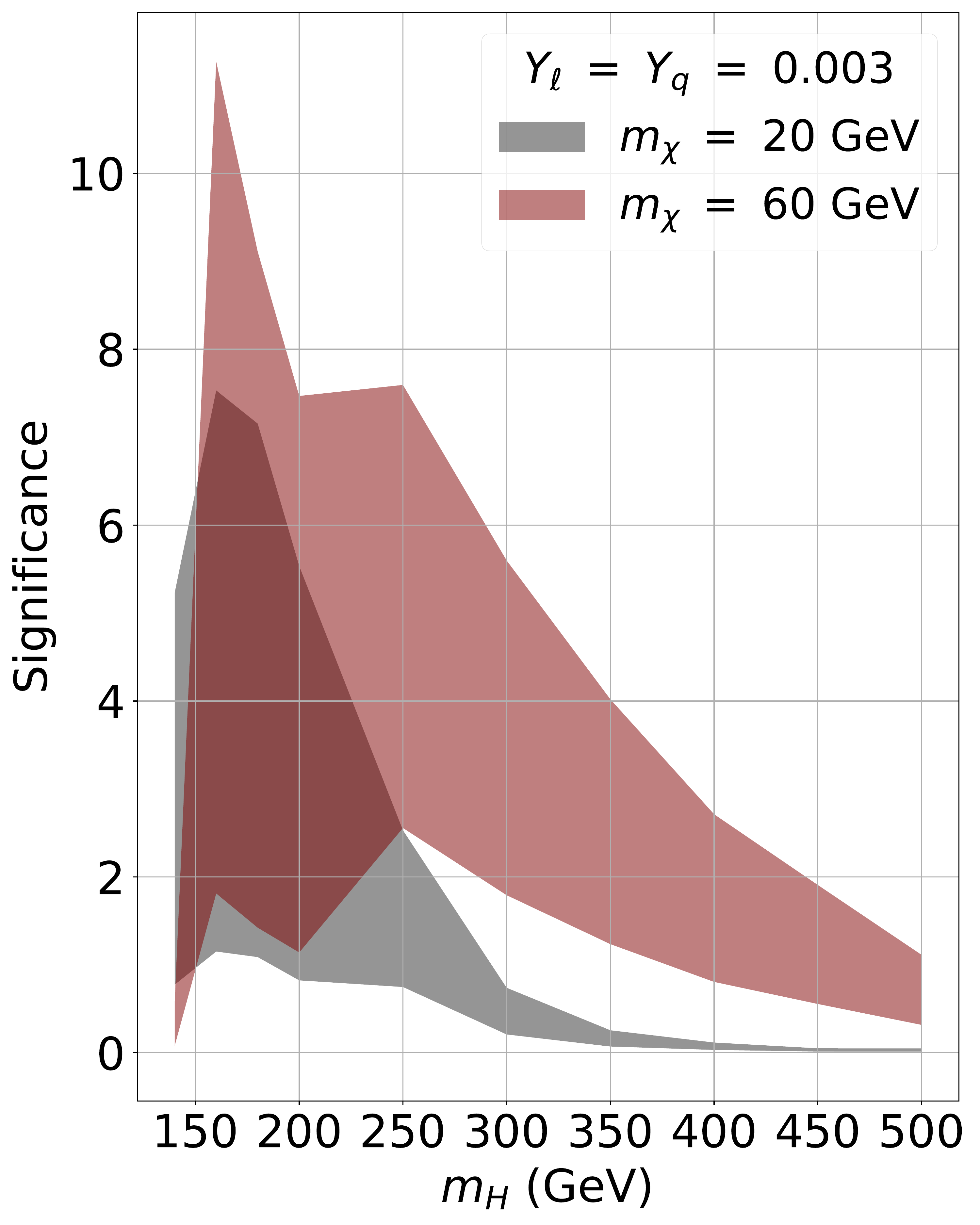}
}
\subfloat[]{
  \label{fig:signif_0p005}
  \centering
  \includegraphics[width=0.3\textwidth]{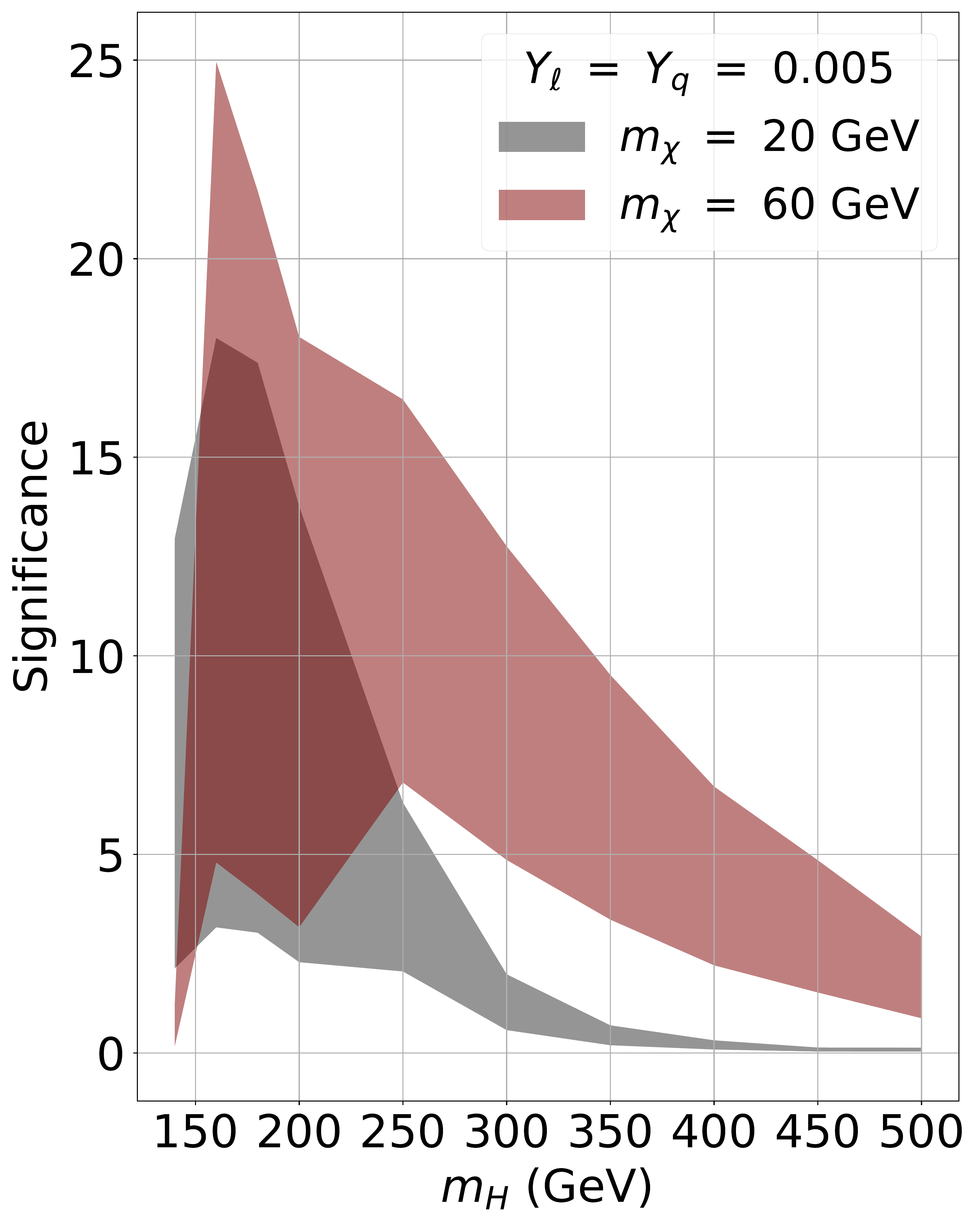}
}
\subfloat[]{
  \label{fig:signif_0p01}
  \centering
  \includegraphics[width=0.3\textwidth]{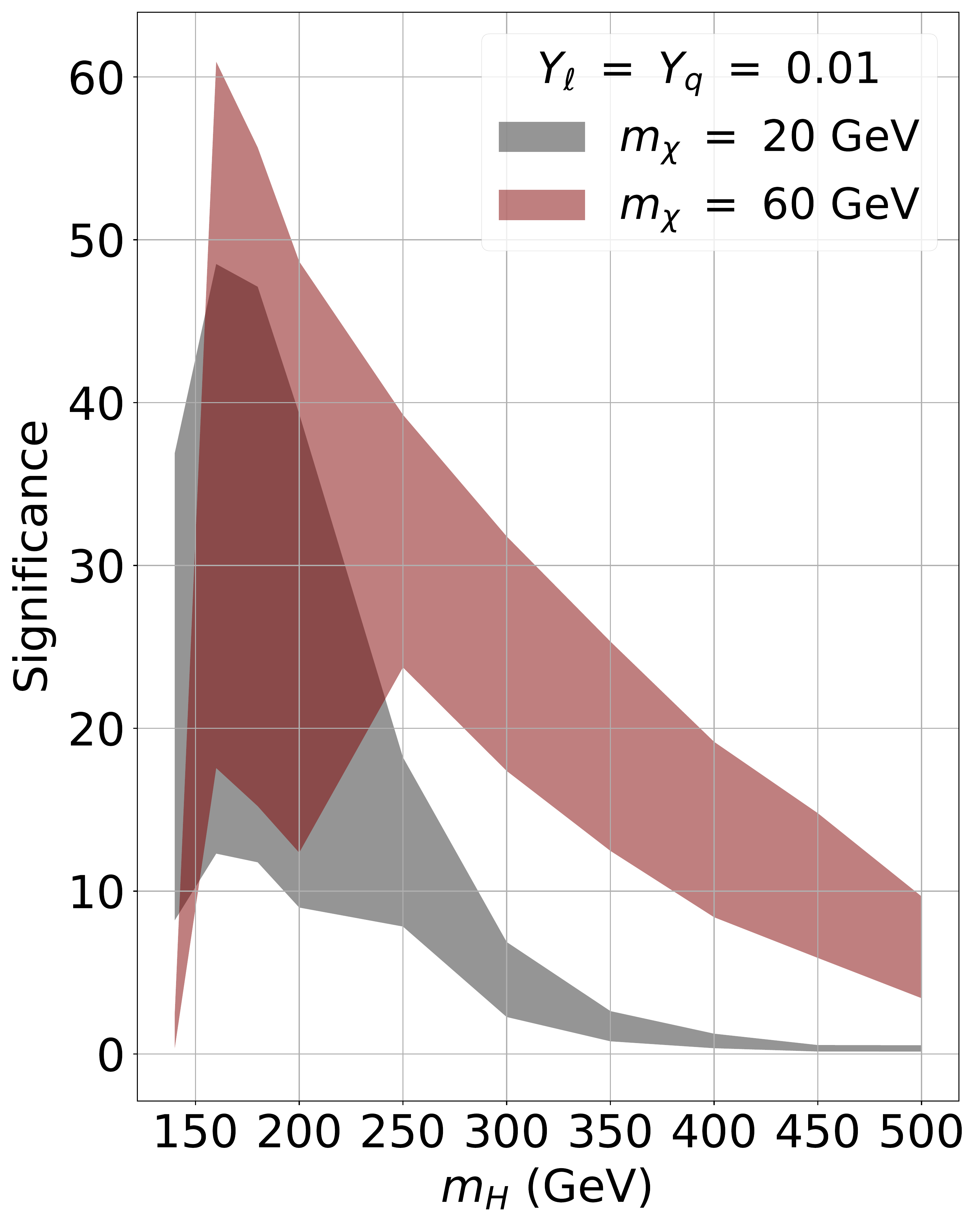}
}
\caption{\small \em Variation of significance with $m_H$ for $m_\chi\,=\,20,\,60$ GeV and $Y_\ell\,=\,Y_q$
  (a) $0.003$ (b) $0.005$ and (c) $0.01$ by scanning $\alpha$ in the range $(0-0.2)$
  for ${\cal L}\,=\,1000\,fb^{-1}$. For each colored shade, the outer and inner edges correspond to $\alpha\,=\,0$ and $0.2$, respectively.}
\label{fig:signif_ext}
\end{figure}
can be applied regardless of the values of $m_\chi$ in the given range, while for relatively lighter $m_H$ different values of
$m_\chi$ require different $\Delta R$ cuts. The boost of the decay products from $H$ is the
crucial factor here. For $m_H\,\geq\,250$ GeV, we therefore apply a  uniform cut $\Delta R\,\leq\,2$
(as the best possible choice) to improve the signal significance.

\begin{table}[h!]
  \begin{center}
    {\footnotesize
    \begin{tabular}{|l|c|c|c|}
      \hline Significance $({\cal S})$ & \multicolumn{3}{c|}{Systematic uncertainty $=\,0\,/\,10\,/\,20\,(\%)$}  \\
      \hline\hline
      \diagbox{$(m_\chi, m_H)$ (GeV)}{$(Y_\ell, Y_q)$} & $(0.003,0.003)$ & $(0.005,0.005)$ & $(0.01,0.01)$  \\
      \hline
      \texttt{$(20,160)$}                        & $7.53/2.22/1.15$    & $18.01/6.05/3.16$   & $48.51/22.53/12.31$ \\
      \texttt{$(20,250)$}                        & $2.53/1.33/0.75$    & $6.31/3.58/2.05$    & $18.22/12.56/7.83$  \\
      \texttt{$(20,400)$}                        & $0.12/0.06/0.03$    & $0.32/0.16/0.09$    & $1.25/0.64/0.36$    \\
      \hline
      \texttt{$(40,160)$}                        & $8.79/2.64/1.37$    & $20.61/7.15/3.74$   & $53.79/26.17/14.43$ \\
      \texttt{$(40,250)$}                        & $8.07/4.74/2.75$    & $17.63/11.89/7.37$  & $41.18/34.77/25.48$ \\
      \texttt{$(40,400)$}                        & $1.51/0.77/0.43$    & $3.91/2.11/1.99$    & $12.33/7.79/4.56$  \\
      \hline
      \texttt{$(60,160)$}                        & $11.26/3.49/1.81$   & $24.95/9.12/4.79$    & $60.94/31.42/17.56$  \\
      \texttt{$(60,250)$}                        & $7.59/4.42/2.56$    & $16.45/11.07/6.81$   & $39.26/32.79/23.74$  \\
      \texttt{$(60,400)$}                        & $2.71/1.43/0.81$    & $6.71/3.84/2.2.1$    & $19.17/13.38/8.41$   \\
      \hline
    \end{tabular}}
  \end{center}
  \caption{\small \em Signal significance for different benchmark scenarios for ${\cal L}\,=\,1000\,fb^{-1}$
    with $0\%$, $10\%$ and $20\%$ systematic uncertainties on background contribution.}
  \label{tab:signif_ext}
\end{table}

We now show the impact of including the background only systematic uncertainties by introducing a parameter ($\alpha$)
in the modified expression of signal significance ${\cal S}\,=\,S\,/\,\sqrt{S + B + (\alpha B)^2}$ \cite{WinNT}.
In Figure \ref{fig:signif_ext}, we show the drop in signal significance by varying $\alpha$ in the range ($0-0.2$),
and in Table \ref{tab:signif_ext}, we display the effects of systematics for a few representative
benchmark points. While the significance drops with increasing systematics,
for some parameter choices the situation still remains quite promising even after including a $(10-20)\%$ systematic uncertainties.

\section{Summary and outlook} \label{Summary}

Any hypothetical Higgs-like state with SM-like couplings with ordinary
matter and gauge fields would incur a serious constraint from its non
observation at the LHC. What happens if the couplings are
unconventional?  In this analysis we explored whether rather lght,
${\cal{O}}$\,($10\,-\,100$) GeV exotic CP-even ($H$) and CP-odd
($\chi$) states having unusual Yukawa and gauge couplings could be
observed through a resonance signature at the 14 TeV run of the LHC
initially with 300 ${\rm fb}^{-1}$ data, where there is a possibility
for a 10-fold increase in luminosity (HL-LHC). Their Yukawa couplings
are purely flavor off-diagonal, $H(\chi)ff'$, and there is no $HVV$
(where $V\equiv Z,W$) interaction. To emphasize that our working
hypothesis is not based on unrealizable wild assumptions, we point out
that a broad class of flavor models based on discrete symmetry groups
does predict such properties for nonstandard spin-$0$ states.  Just
with these two generic properties, these exotic states can avoid the
conventional experimental constraints on their masses.  Admittedly, to
bypass the stringent constraints from $D$ meson mixing an approximate
adjustment between $Y_q^H$ and $Y_q^\chi$ had to be innovated to yield
results that are experimentally exciting.  We have assumed $m_\chi \ll
m_H$.
We performed a detailed study with a few set of benchmark points :
$m_H\,=\,(160,\,170)$, $m_\chi\,=\,(20,60)$, all in GeV. Subsequently,
we include a wider range of benchmark points and incorporated possible
systematic uncertainties to examine the expected signal significance. 
We displayed the significance as a function of those off-diagonal Yukawa couplings with
quarks and leptons. Although the ancestral origin of these peculiar
couplings may be traced, as mentioned before, in discrete flavor
symmetry models, we remained agnostic about the UV picture of the new
physics, and took a model independent view relying only on the overall
gross pattern without committing to model-specific values of those
Yukawa couplings.

The wisdom we gather from our analysis is that in general the
multivariate analysis performs better than the traditional cut based
method.  As expected, larger Yukawa values show higher signal
significance, and chances exist that the relatively larger couplings
we chose as benchmark values, otherwise allowed by existing indirect
constraints, can be ruled in or out with 300 ${\rm fb}^{-1}$ data at
14 TeV. Some of those couplings can even be tested at the upcoming
13.6 TeV Run 3 data. In fact, even in the existing 13 TeV Run 2 data,
such exotic treasures might be hidden which require an in-depth
experimental analysis for manifestation.  To explore the smaller
values of those couplings we cannot but rely on the HL-LHC.
We admit that our analysis does not contain the following issues that
experimental groups do pay attention to : jet faking as $\tau_h$
and/or leptons, lepton charge misidentification, photon conversions
into lepton pairs, uncertainties on luminosity and trigger
efficiencies, etc \cite{CMS:2018jrd}. However, we have considered a
linear-in-background systematic uncertainty to probe the signal sensitivity at the HL-LHC.
Still, our results look quite promising and we hope that the initial steps we have taken in
this paper would encourage the ATLAS and CMS exotica groups to pursue
further in this direction.

\section*{Acknowledgement}

We thank S. Bhattacharya, T. Jha and I. Chakraborty for
discussions. We acknowledge the central computing
facility of Saha Institute of Nuclear Physics for computational
support.

\bibliographystyle{JHEP}
\typeout{}
\bibliography{exotic_higgs.bib}

\end{document}